\documentstyle[11pt,aaspp4]{article}
\begin{document}

\title{DIRECT Distances to Nearby Galaxies Using Detached Eclipsing
Binaries and Cepheids. IV. Variables in the Field M31D\footnote{Based
on the observations collected at the Michigan-Dartmouth-MIT (MDM)
1.3~m telescope and at the F.~L.~Whipple Observatory (FLWO)
1.2~m telescope}}

\author{J. Kaluzny and B. J. Mochejska}
\affil{Warsaw University Observatory, Al. Ujazdowskie 4,
PL-00-478 Warszawa, Poland} 
\affil{\tt e-mail: jka@sirius.astrouw.edu.pl, mochejsk@sirius.astrouw.edu.pl} 
\author{K. Z. Stanek\altaffilmark{2}, M. Krockenberger and D. D. Sasselov } 
\affil{Harvard-Smithsonian Center for Astrophysics, 60 Garden St.,
Cambridge, MA~02138} 
\affil{\tt e-mail: kstanek@cfa.harvard.edu, krocken@cfa.harvard.edu, 
sasselov@cfa.harvard.edu} 
\altaffiltext{2}{On leave from N.~Copernicus Astronomical Center, 
Bartycka 18, Warszawa PL-00-716, Poland} 
\author{J. L. Tonry} 
\affil{University of Hawaii, Institute for Astronomy, 
2680 Woodlawn Dr., Honolulu, HI~96822}
\affil{\tt e-mail: jt@avidya.ifa.hawaii.edu} 
\author{M. Mateo}
\affil{Department of Astronomy, University of Michigan, 
821 Dennison Bldg., Ann Arbor, MI~48109--1090} 
\affil{\tt e-mail: mateo@astro.lsa.umich.edu}

\begin{abstract}

We undertook a long term project, DIRECT, to obtain the direct
distances to two important galaxies in the cosmological distance
ladder -- M31 and M33 -- using detached eclipsing binaries (DEBs) and
Cepheids. While rare and difficult to detect, DEBs provide us with the
potential to determine these distances with an accuracy better than
5\%. The extensive photometry obtained in order to detect DEBs
provides us with good light curves for the Cepheid variables. These
are essential to the parallel project to derive direct Baade-Wesselink
distances to Cepheids in M31 and M33. For both Cepheids and eclipsing
binaries, the distance estimates will be free of any intermediate
steps.
 
As a first step in the DIRECT project, between September 1996 and
October 1997 we obtained 95 full/partial nights on the F. L. Whipple
Observatory 1.2~m telescope and 36 full nights on the
Michigan-Dartmouth-MIT 1.3~m telescope to search for DEBs and new
Cepheids in the M31 and M33 galaxies.  In this paper, fourth in the
series, we present the catalog of variable stars, most of them newly
detected, found in the field M31D $[(\alpha,\delta)=
(11.\!\!\arcdeg03, 41.\!\!\arcdeg27), {\rm J2000.0}]$.  We have found
71 variable stars: 5 eclipsing binaries, 38 Cepheids and 28 other
periodic, possible long period or non-periodic variables. The catalog
of variables, as well as their photometry and finding charts, is
available via {\tt anonymous ftp} and the {\tt World Wide Web}.  The
complete set of the CCD frames is available upon request.

\end{abstract}


\section{Introduction}

The two nearby galaxies M31 and M33 are stepping stones to most of our
current effort to understand the evolving universe at large scales.
First, they are essential to the calibration of the extragalactic
distance scale (Jacoby et al.~1992; Tonry et al.~1997). Second, they
constrain population synthesis models for early galaxy formation and
evolution and provide the stellar luminosity calibration. There is one
simple requirement for all this---accurate distances.
 
Detached eclipsing binaries (DEBs) have the potential to establish
distances to M31 and M33 with an unprecedented accuracy of better than
5\% and possibly to better than 1\%. These distances are now known to
no better than 10-15\%, as there are discrepancies of $0.2-0.3\;{\rm
mag}$ between various distance indicators (e.g.~Huterer, Sasselov \&
Schechter 1995; Holland 1998; Stanek \& Garnavich 1998).  Detached
eclipsing binaries (for reviews see Andersen 1991; Paczy\'nski 1997)
offer a single step distance determination to nearby galaxies and may
therefore provide an accurate zero point calibration---a major step
towards very accurate determination of the Hubble constant, presently
an important but daunting problem for astrophysicists. A DEB system
was recently used by Guinan et al.~(1998) and Udalski et al.~(1998) to
obtain an accurate distance estimate to the Large Magellanic Cloud.
 
The detached eclipsing binaries have yet to be used (Huterer et
al.~1995; Hilditch 1996) as distance indicators to M31 and M33.
According to Hilditch (1996), there were about 60 eclipsing binaries
of all kinds known in M31 (Gaposchkin 1962; Baade \& Swope 1963, 1965)
and only {\em one} in M33 (Hubble 1929).  Only now does the
availability of large-format CCD detectors and inexpensive CPUs make
it possible to organize a massive search for periodic variables, which
will produce a handful of good DEB candidates. These can then be
spectroscopically followed-up with the powerful new 6.5-10 meter
telescopes.

The study of Cepheids in M31 and M33 has a venerable history (Hubble
1926, 1929; Gaposchkin 1962; Baade \& Swope 1963, 1965). In the 1980s,
Freedman \& Madore (1990) and Freedman, Wilson, \& Madore (1991)
studied small samples of the earlier discovered Cepheids, to build
period-luminosity relations in M31 and M33, respectively.  However,
both the sparse photometry and the small samples do not provide a good
basis for obtaining direct Baade-Wesselink distances (see, e.g.,
Krockenberger, Sasselov \& Noyes 1997) to Cepheids---the need for new
digital photometry has been long overdue. More recently, Magnier et
al.~(1997) surveyed large portions of M31, which have previously been
ignored, and found some 130 new Cepheid variable candidates.  Their
light curves are, however, rather sparsely sampled and in the $V$-band
only.

In Kaluzny et al.~(1998, hereafter: Paper I) and Stanek et al.~(1998,
1999, hereafter: Papers II and III), the first three papers of the
series, we presented the catalogs of variable stars found in three
fields in M31, called M31B, M31A and M31C. Here we present the catalog
of variables from the next field M31D. In Sec.2 we discuss the
selection of the fields in M31 and the observations. In Sec.3 we
describe the data reduction and calibration. In Sec.4 we discuss
briefly the automatic selection we used for finding the variable
stars. In Sec.5 we discuss the classification of the variables.  In
Sec.6 we present the catalog of variable stars, followed by brief
discussion of the results in Sec.7.

\section{Fields selection and observations}

M31 was primarily observed with the 1.3~m McGraw-Hill Telescope at the
Michigan-Dartmouth-MIT (MDM) Observatory. We used the
front-illuminated, Loral $2048^2$ CCD ``Wilbur'' (Metzger, Tonry \&
Luppino 1993), which at the $f/7.5$ station of the 1.3~m telescope has
a pixel scale of $0.32\; arcsec\; pixel^{-1}$ and field of view of
roughly $11\;arcmin$. We used Kitt Peak Johnson-Cousins $BVI$ filters.
Data for M31 were also obtained, mostly in 1997, with the 1.2~m
telescope at the F. L. Whipple Observatory (FLWO), where we used
``AndyCam'' (Szentgyorgyi et al.~1999), with a thinned, back-side
illuminated, AR coated Loral $2048^2$ pixel CCD.  The pixel scale
happens to be essentially the same as at the MDM 1.3~m telescope. We
used standard Johnson-Cousins $BVI$ filters.

Fields in M31 were selected using the MIT photometric survey of M31 by
Magnier et al.~(1992) and Haiman et al.~(1994) (see Paper I, Fig.1).
We selected six $11'\times11'$ fields, M31A--F, four of them (A--D)
concentrated on the rich spiral arm in the northeast part of M31, one
(E) coinciding with the region of M31 searched for microlensing by
Crotts \& Tomaney (1996), and one (F) containing the giant star
formation region known as NGC206 (observed by Baade \& Swope
1963). Fields A--C were observed during September and October 1996
five to eight times per night in the $V$ band, resulting in total of
110--160 $V$ exposures per field. Fields D--F were observed once a
night in the $V$-band. Some exposures in $B$ and $I$ were also
taken. M31 was also observed, in 1996 and 1997, at the FLWO 1.2~m
telescope, whose main target was M33.

In this paper we present the results for the M31D field.  We obtained
for this field useful data during 23 nights at the MDM, collecting a
total of $24\times 900\;sec$ exposures in $V$ and $2\times 600\;sec$
exposures in $I$. We also obtained for this field useful data during
35 nights at the FLWO, in 1996 and 1997, collecting a total of
$67\times 900\;sec$ exposures in $V$, $43\times 600\;sec$ exposures in
$I$ and $5\times 1200\;sec$ exposures of $B$.\footnote{The complete
list of exposures for this field and related data files are available
through {\tt anonymous ftp} on {\tt cfa-ftp.harvard.edu}, in {\tt
pub/kstanek/DIRECT} directory. Please retrieve the {\tt README} file
for instructions.  Additional information on the DIRECT project is
available through the {\tt WWW} at {\tt
http://cfa-www.harvard.edu/\~\/kstanek/DIRECT/}.}

\section{Data reduction, calibration and astrometry}

The details of the reduction procedure were given in Paper I.
Preliminary processing of the CCD frames was done with the standard
routines in the IRAF-CCDPROC package.\footnote{IRAF is distributed by
the National Optical Astronomy Observatories, which are operated by
the Associations of Universities for Research in Astronomy, Inc.,
under cooperative agreement with the NSF} Stellar profile photometry
was extracted using the {\it Daophot/Allstar} package (Stetson 1987,
1992).  We selected a ``template'' frame for each filter using a
single frame of particularly good quality.  These template images were
reduced in a standard way (Paper I).  Other images were reduced using
{\it Allstar} in the fixed-position-mode using as an input the
transformed object list from the template frames.  For each frame the
list of instrumental photometry derived for a given frame was
transformed to the common instrumental system of the appropriate
``template'' image.  Photometry obtained for the $B,V$ and $I$ filters
was combined into separate data bases. M31D images obtained at the
FLWO were reduced using MDM ``templates''.  In case of $B$-band images
obtained at FLWO we used the $V$-band MDM template to fix the
positions of the stars.

The photometric $VI$ calibration of the MDM data was discussed in
Paper I. In addition, for the field M31D on the night of 1997 October
9/10 we have obtained independent $BVI$ calibration with the FLWO
1.2~m telescope. There was an offset of $-0.014\;{\rm mag}$ in $V$ and
$0.047\;{\rm mag}$ in $V-I$ between the FLWO and the MDM calibration,
i.e. within our estimate of the total $0.05\;mag$ systematic error
discussed in Paper I.

To check the internal consistency of our photometry we compared the
photometry for 55 $V<20$ and 100 $I<20$ common stars in the overlap
region between the fields M31C and M31D (Figure~\ref{fig:xy}).  There
was an offset of $-0.063\;{\rm mag}$ in $V$, $-0.040\; {\rm mag}$ in
$I$ and $0.007\;{\rm mag}$ in $B$.  We also derived equatorial
coordinates for all objects included in the data bases for the $V$
filter. The transformation from rectangular coordinates to equatorial
coordinates was derived using 196 stars identified in the list
published by Magnier et al.~(1992).

\section{Selection of variables}

\begin{figure}[t]
\plotfiddle{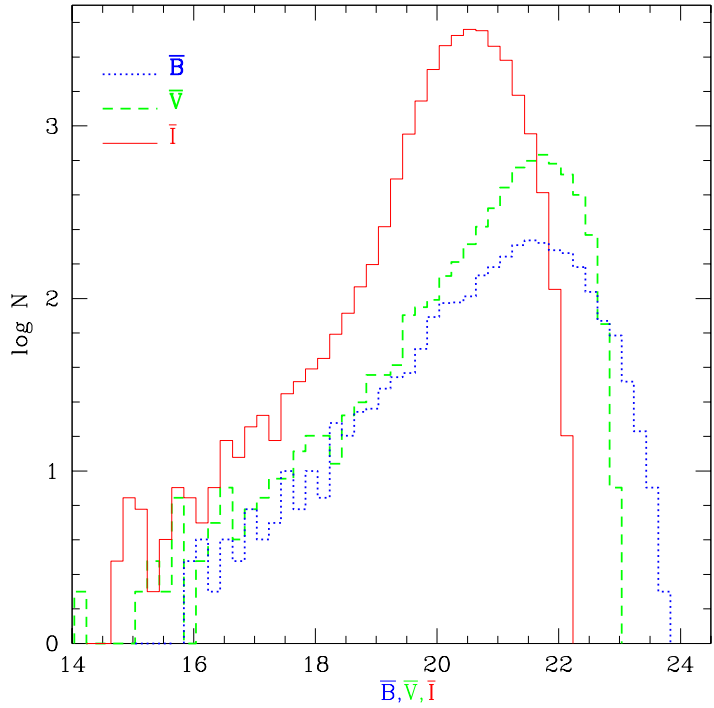}{8cm}{0}{50}{50}{-160}{-85}
\caption{Distributions in $B$ (dotted line), $V$ (dashed line) and $I$
(continuous line) of stars in the field M31D.}
\label{fig:dist}
\end{figure}

The procedure for selecting the variables was described in detail in
Paper I, so here we only give a short description, noting changes when
necessary.  The reduction procedure described in previous section
produces databases of calibrated $BVI$ magnitudes and their standard
errors. The $BV$ databases for M31D field contain 8016 stars, with up 
to 91 measurements in $V$ and up to 5 measurements in $B$, and the 
$I$ database contains 35576 stars with up to 45 measurements.
Figure~\ref{fig:dist} shows the distributions of stars as a function of
mean $\bar{B}$, $\bar{V}$ or $\bar{I}$ magnitude.  As can be seen from
the shape of the histograms, our completeness starts to drop rapidly
at about $\bar{B}\sim23$, $\bar{V}\sim22$ and $\bar{I}\sim20.5$. The
primary reason for this difference in the depth of the photometry
between $BV$ and $I$ is the level of the combined sky and background
light, which is about three times higher in the $I$ filter than in the
$BV$ filters.

The measurements flagged as ``bad'' and measurements with errors
exceeding the average error by more than $4\sigma$ are removed
(Paper~I).  Usually zero to 5 points are removed, leaving the
majority of stars with roughly $N_{good}\sim85-90$ $V$\/
measurements.  For further analysis we use only those stars that have
at least $N_{good}>N_{max}/2\;(=45)$ measurements. There are 5813
such stars in the $V$ database of the M31D field.

\begin{figure}[t]
\plotfiddle{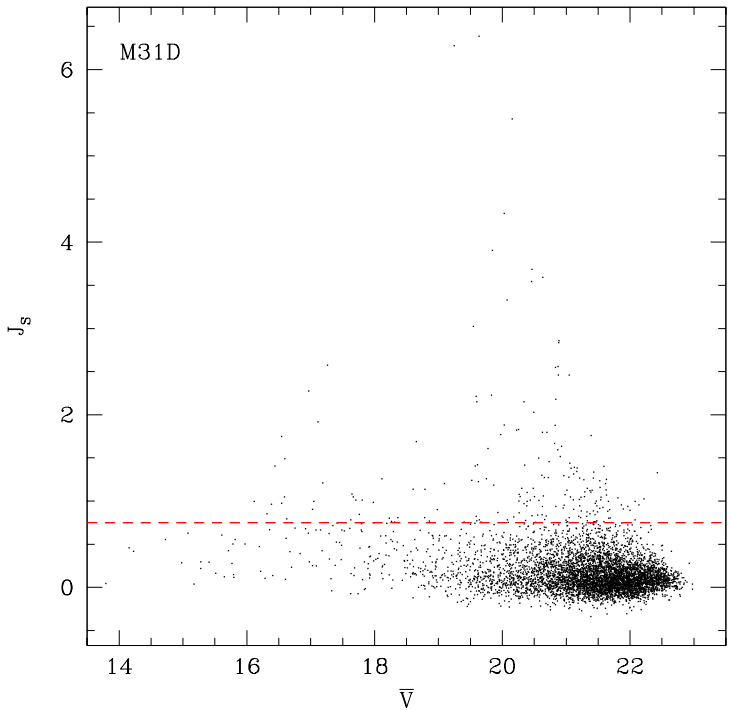}{8cm}{0}{50}{50}{-160}{-85}
\caption{Variability index $J_S$ vs. mean $\bar{V}$ magnitude for 5813
stars in the field M31D with $N_{good}>45$.  Dashed line at $J_S=0.75$
defines the cutoff applied for variability.}
\label{fig:stetj}
\end{figure}

Our next goal is to select objectively a sample of variable stars from
the total sample defined above.  There are many ways to proceed, and
we largely follow the approach of Stetson (1996).  The procedure is
described in more detail in Paper~I. In short, for each star we
compute the Stetson's variability index $J_S$ (Paper I, Eq.7), and
stars with values exceeding some minimum value $J_{S,min}$ are
considered candidate variables.  The definition of Stetson's
variability index includes the standard errors of individual
observations.  If, for some reason, these errors were over- or
underestimated, we would either miss real variables, or select
spurious variables as real ones. Using the procedure described in
Paper I, we scale the {\em Daophot} errors to better represent the
``true'' photometric errors.  We then select the candidate variable
stars by computing the value of $J_S$ for the stars in our $V$
database.  We used a cutoff of $J_{S,min}=0.75$ and additional cuts
described in Paper I to select 178 candidate variable stars (about 3\%
of the total number of 5813).  In Figure~\ref{fig:stetj} we plot the
variability index $J_S$ vs. apparent visual magnitude $\bar{V}$ for
5813 stars with $N_{good}>45$.

\section{Period determination, classification of variables}

We based our candidate variables selection on the $V$ band data
collected at the MDM and the FLWO telescopes.  We also have the
$BI$-bands data for the field, up to 45 $I$-band epochs and up to 5
$B$-band epochs, although for a variety of reasons some of the
candidate variable stars do not have an $B$ or $I$-band
counterpart. We will therefore not use the $BI$ data for the period
determination and broad classification of the variables. We will
however use the $BI$ data for the ``final'' classification of some
variables.

Next we searched for the periodicities for all 178 candidate
variables, using a variant of the Lafler-Kinman (1965) technique
proposed by Stetson (1996). Starting with the minimum period of
$0.25\;days$, successive trial periods are chosen so
\begin{equation}
P_{j+1}^{-1}=P_{j}^{-1}-\frac{0.02}{\Delta t},
\end{equation}
where $\Delta t=t_{N}-t_{1}=398\;days$ is the time span of the series.
The maximum period considered is $150\;days$.  For each candidate
variable 10 best trial periods are selected (Paper I) and then used in
our classification scheme.

The variables we are most interested in are Cepheids and eclipsing
binaries (EBs). We therefore searched our sample of variable stars for
these two classes of variables. As mentioned before, for the broad
classification of variables we restricted ourselves to the $V$ band
data.  We will, however, present and use the $BI$-bands data, when
available, when discussing some of the individual variable stars.

For EBs we used search strategy described in Paper II.  Within our
assumption the light curve of an EB is determined by nine parameters:
the period, the zero point of the phase, the eccentricity, the
longitude of periastron, the radii of the two stars relative to the
binary separation, the inclination angle, the fraction of light coming
from the bigger star and the uneclipsed magnitude. A total of seven
variables passed all of the criteria. We then went back to the CCD
frames and tried to see by eye if the inferred variability is indeed
there, especially in cases when the light curve is very noisy/chaotic.
We decided to remove two dubious eclipsing binaries.  The remaining
five EBs with their parameters and light curves are presented in the
Section~6.1.

In the search for Cepheids we followed the approach by Stetson (1996)
of fitting template light curves to the data. We used the
parameterization of Cepheid light curves in the $V$-band as given by
Stetson (1996). There was a total of 65 variables passing all of the
criteria (Paper I and II), but after investigating the CCD frames we
removed 12 dubious ``Cepheids'', which leaves us with 53 probable
Cepheids. Their parameters and light curves are presented in the
Sections~6.2,~6.3.

After the preliminary selection of seven eclipsing binaries and 65
possible Cepheids, we were left with 106 ``other'' variable stars.
After raising the threshold of the variability index to
$J_{S,min}=1.2$ (Paper I) we are left with 25 variables. After
investigating the CCD frames we removed 12 dubious variables from the
sample, which leaves 13 variables which we classify as
miscellaneous. Their parameters and light curves are presented in the
Section~6.4.

\section{Catalog of variables}

In this section we present light curves and some discussion of the 71
variable stars discovered by our survey in the field M31D.
\footnote{Complete $V$ and (when available) $BI$ photometry and
$128\times128\;pixel$ ($\sim 40''\times40''$) $V$ finding charts for
all variables are available from the authors via the {\tt anonymous
ftp} from the Harvard-Smithsonian Center for Astrophysics and can be
also accessed through the {\tt World Wide Web}.}  The variable stars
are named according to the following convention: letter V for
``variable'', the number of the star in the $V$ database, then the
letter ``D'' for our project, DIRECT, followed by the name of the
field, in this case (M)31D, e.g. V1599 D31D.  Tables~\ref{table:ecl},
\ref{table:ceph}, \ref{table:per} and \ref{table:misc} list the
variable stars sorted broadly by four categories: eclipsing binaries,
Cepheids, other periodic variables and ``miscellaneous'' variables, in
our case meaning ``variables with no clear periodicity''. Some of the
variables which were found independently by survey of Magnier et
al.~(1997) are denoted in the ``Comments'' by ``Ma97 ID'', where the
``ID'' is the identification number assigned by Magnier at
al.~(1997). We also cross-identify several variables found by us in
Paper III.

\subsection{Eclipsing binaries}

In Table~\ref{table:ecl} we present the parameters of the 5 eclipsing
binaries in the M31D field.  The lightcurves of these variables are
shown in Figure~\ref{fig:ecl}, along with the simple eclipsing binary
models discussed in the Paper I.  The variables are sorted in the
Table~\ref{table:ecl} by the increasing value of the period $P$. For
each eclipsing binary we present its name, J2000.0 coordinates (in
degrees), period $P$, magnitudes $V_{max}, I_{max}$ and $B_{max}$ of
the system outside of the eclipse, and the radii of the binary
components $R_1,\;R_2$ in the units of the orbital separation.  We
also give the inclination angle of the binary orbit to the line of
sight $i$ and the eccentricity of the orbit $e$. The reader should
bear in mind that the values of $V_{max},\;I_{max},\;B_{max},\;
R_1,\;R_2,\;i$ and $e$ are derived with a straightforward model of the
eclipsing system, so they should be treated only as reasonable
estimates of the ``true'' value.



\begin{figure}[p]
\plotfiddle{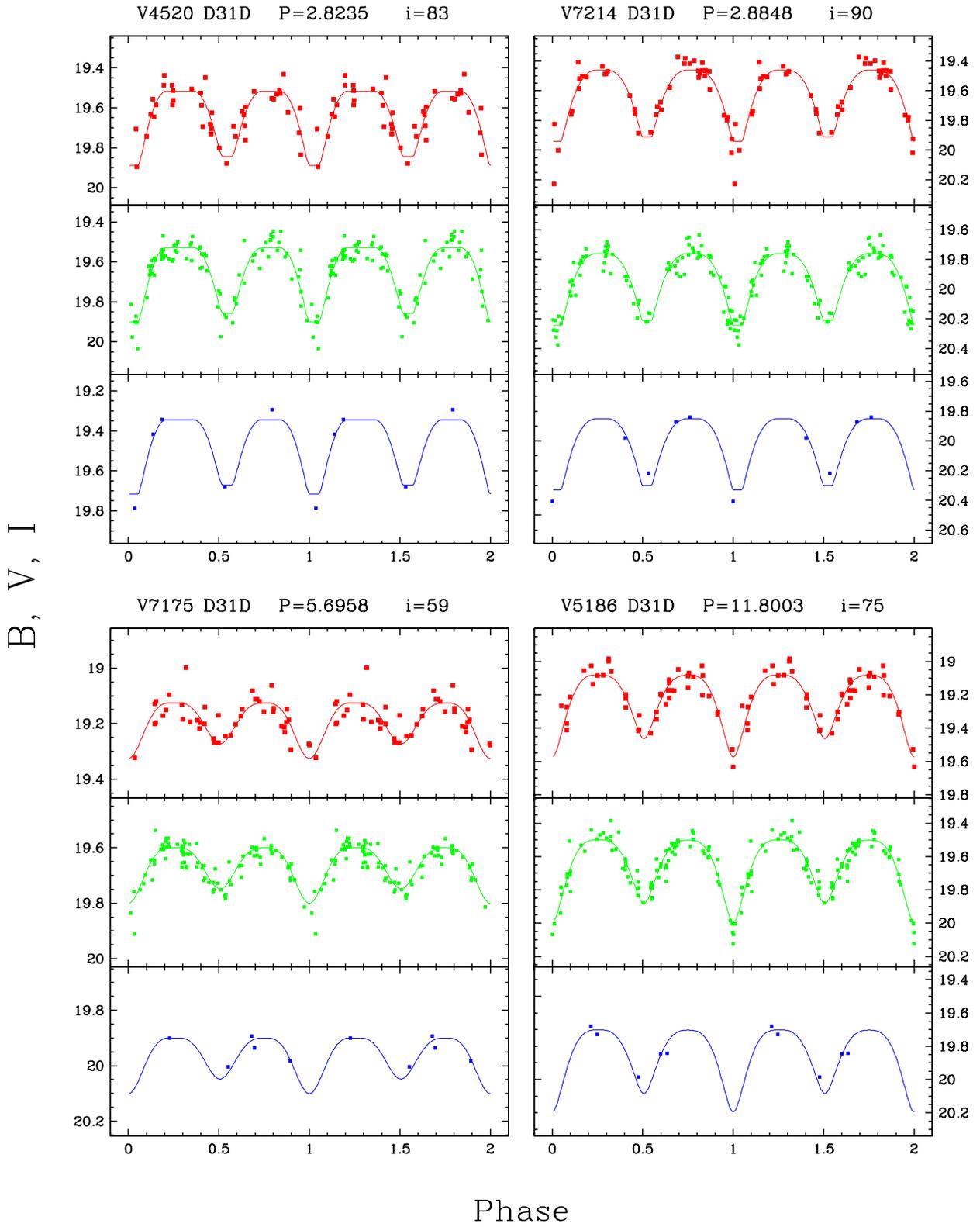}{19.5cm}{0}{83}{83}{-260}{-40}
\caption{$BVI$ lightcurves of eclipsing binaries found in the field
M31D. The thin continuous line represents the best fit model for each
star and photometric band. $B$-band lightcurve is shown in the bottom
panel and $I$-band lightcurve (when present) is shown in the top
panel.}
\label{fig:ecl}
\end{figure}
 
\addtocounter{figure}{-1}
\begin{figure}[t]
\plotfiddle{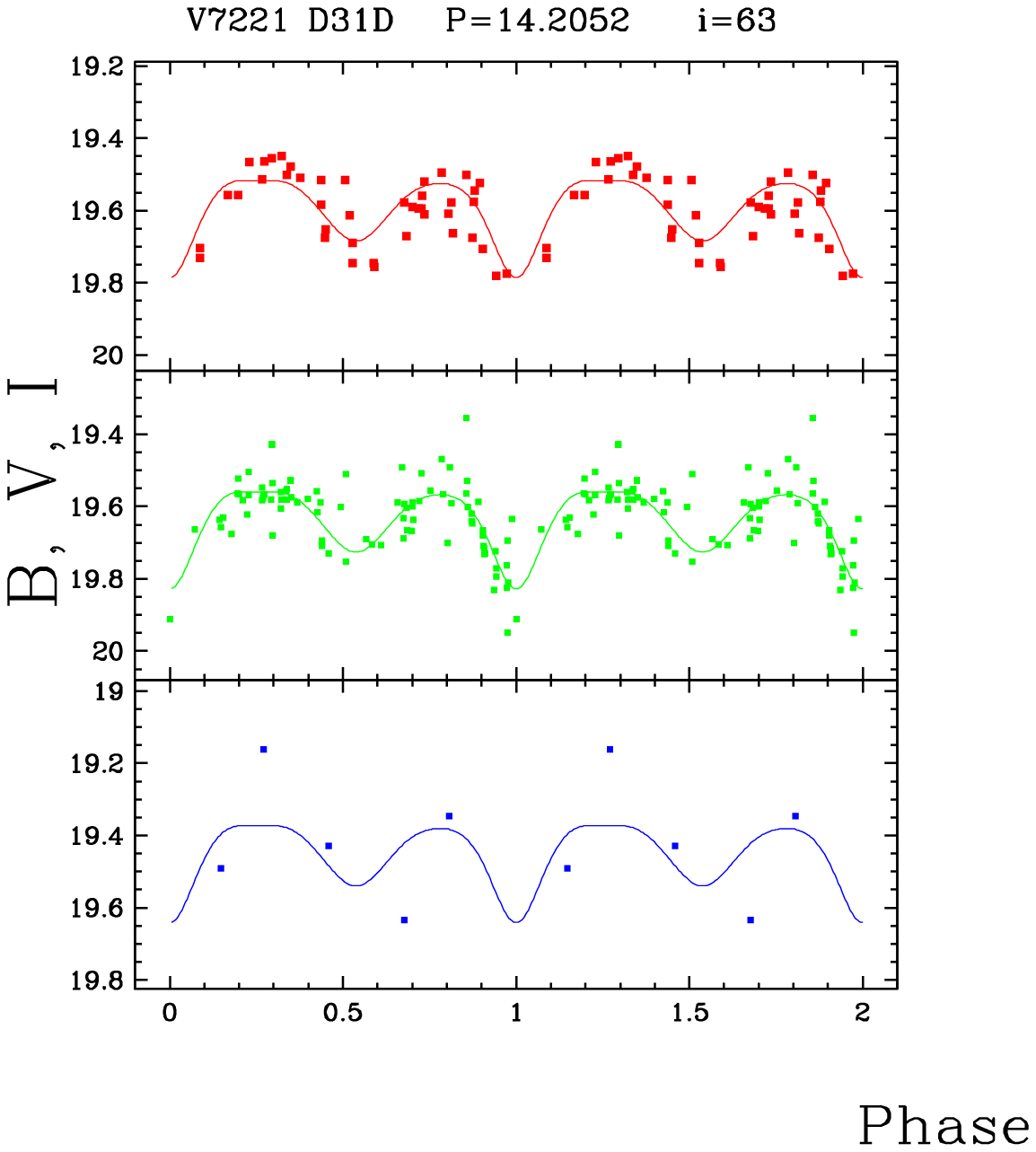}{8cm}{0}{83}{83}{-260}{-360}
\caption{Continued.}
\end{figure}

\begin{figure}[p]
\plotfiddle{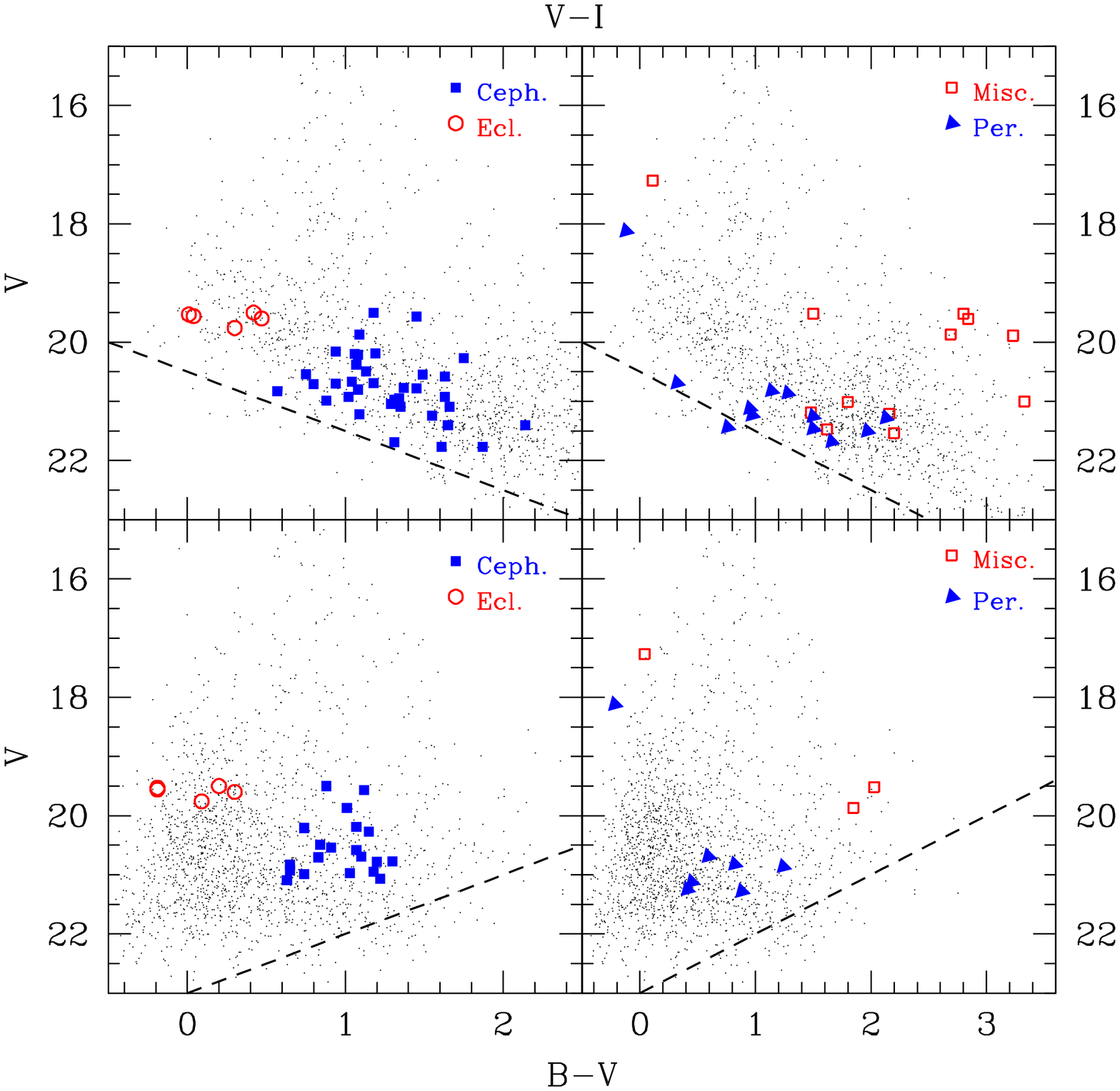}{13cm}{0}{85}{85}{-260}{-135}
\caption{$V,\;V-I$ (upper panels) and $B,\;B-V$ (lower panels)
color-magnitude diagrams for the variable stars found in the field
M31D. The eclipsing binaries and Cepheids are plotted in the left
panels and the other periodic variables and miscellaneous variables
are plotted in the right panels. The dashed lines correspond to the
$I$ detection limit of $I\sim20.5\;{\rm mag}$ (upper panels) and the $B$
detection limit of $B\sim23.\;{\rm mag}$ (lower panels).
\label{fig:cmd}}
\end{figure}

\begin{small}
\tablenum{1} 
\begin{planotable}{lrrrrrrrrcr}
\tablewidth{45pc}
\tablecaption{\sc DIRECT Eclipsing Binaries in M31D}
\tablehead{ \colhead{Name} & \colhead{$\alpha_{J2000.0}$} & \colhead{$\delta_{J2000.0}$} 
& \colhead{$P$} & \colhead{} & \colhead{} & \colhead{} &
\colhead{} & \colhead{} & \colhead{$i$} & \colhead{} \\ 
\colhead{(D31D)} & \colhead{$(\deg)$} & \colhead{$(\deg)$}
& \colhead{$(days)$} & \colhead{$V_{max}$}  & \colhead{$I_{max}$} 
& \colhead{$B_{max}$} & \colhead{$R_1$} & \colhead{$R_2$} & \colhead{(deg)}  
& \colhead{$e$} } 
\startdata
 V4520\ldots   & 11.0001 & 41.3004 &  2.8235 & 19.53 & 19.52 & 19.34 & 0.56 & 0.35 & 83 & 0.03 \nl 
 V7214\dotfill & 11.0840 & 41.2994 &  2.8848 & 19.76 & 19.46 & 19.85 & 0.57 & 0.42 & 90 & 0.00 \nl 
 V7175\dotfill & 11.0824 & 41.2991 &  5.6958 & 19.60 & 19.13 & 19.90 & 0.63 & 0.36 & 59 & 0.01 \nl 
 V5186\dotfill & 11.0158 & 41.2888 & 11.8003 & 19.50 & 19.08 & 19.70 & 0.53 & 0.47 & 75 & 0.01 \nl 
 V7221\dotfill & 11.0845 & 41.2973 & 14.2052 & 19.56 & 19.52 & 19.37 & 0.64 & 0.36 & 63 & 0.14 \nl 
\enddata 

\label{table:ecl}
\end{planotable}
\end{small}

\subsection{Cepheids}

In Table~\ref{table:ceph} we present the parameters of 38 Cepheids in
the M31D field, sorted by the period $P$.  For each Cepheid we present
its name, J2000.0 coordinates, period $P$, flux-weighted mean
magnitudes $\langle V\rangle$ and (when available) $\langle I\rangle$
and $\langle B\rangle$, and the $V$-band amplitude of the variation
$A$.  In Figure~\ref{fig:ceph} we show the phased $B,V,I$ lightcurves
of our Cepheids. Also shown is the best fit template lightcurve
(Stetson 1996), which was fitted to the $V$ data and then for the $I$
data only the zero-point offset was allowed. For the $B$-band data,
lacking the template lightcurve parameterization (Stetson 1996), we
used the $V$-band template, allowing for different zero-points and
amplitudes. With our limited amounts of $B$-band data this approach
produces mostly satisfactory results, but extending the
template-fitting approach of Stetson (1996) to the $B$-band (and
possibly other popular bands) would be most useful.

\begin{figure}[p]
\plotfiddle{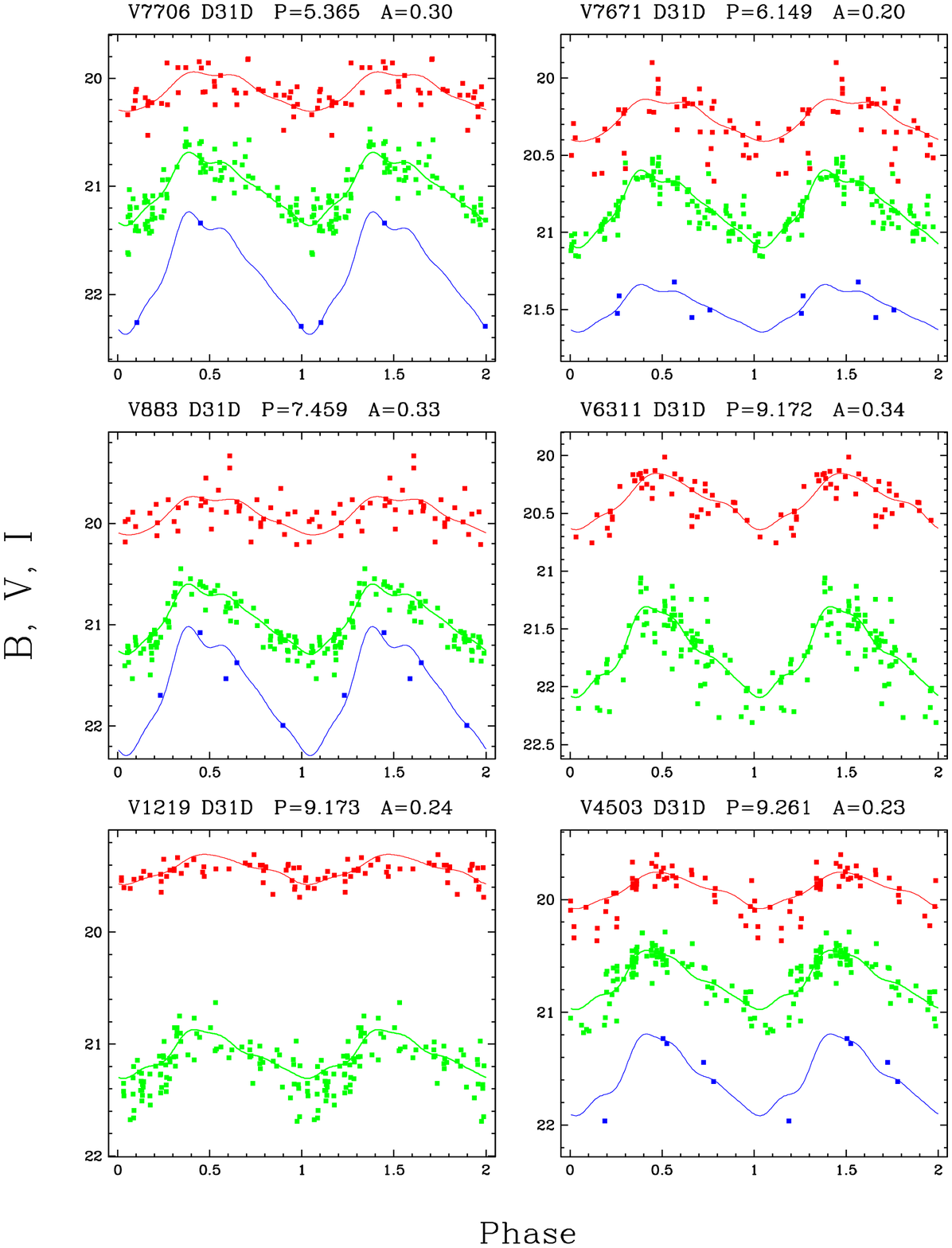}{19.5cm}{0}{83}{83}{-260}{-40}
\caption{$BVI$ lightcurves of Cepheid variables found in the field
M31D. The thin continuous line represents the best fit Cepheid
template for each star and photometric band. $B$ (if present) is
always the faintest and $I$ (if present) is always the brightest.}
\label{fig:ceph}
\end{figure}

\addtocounter{figure}{-1}
\begin{figure}[p]
\plotfiddle{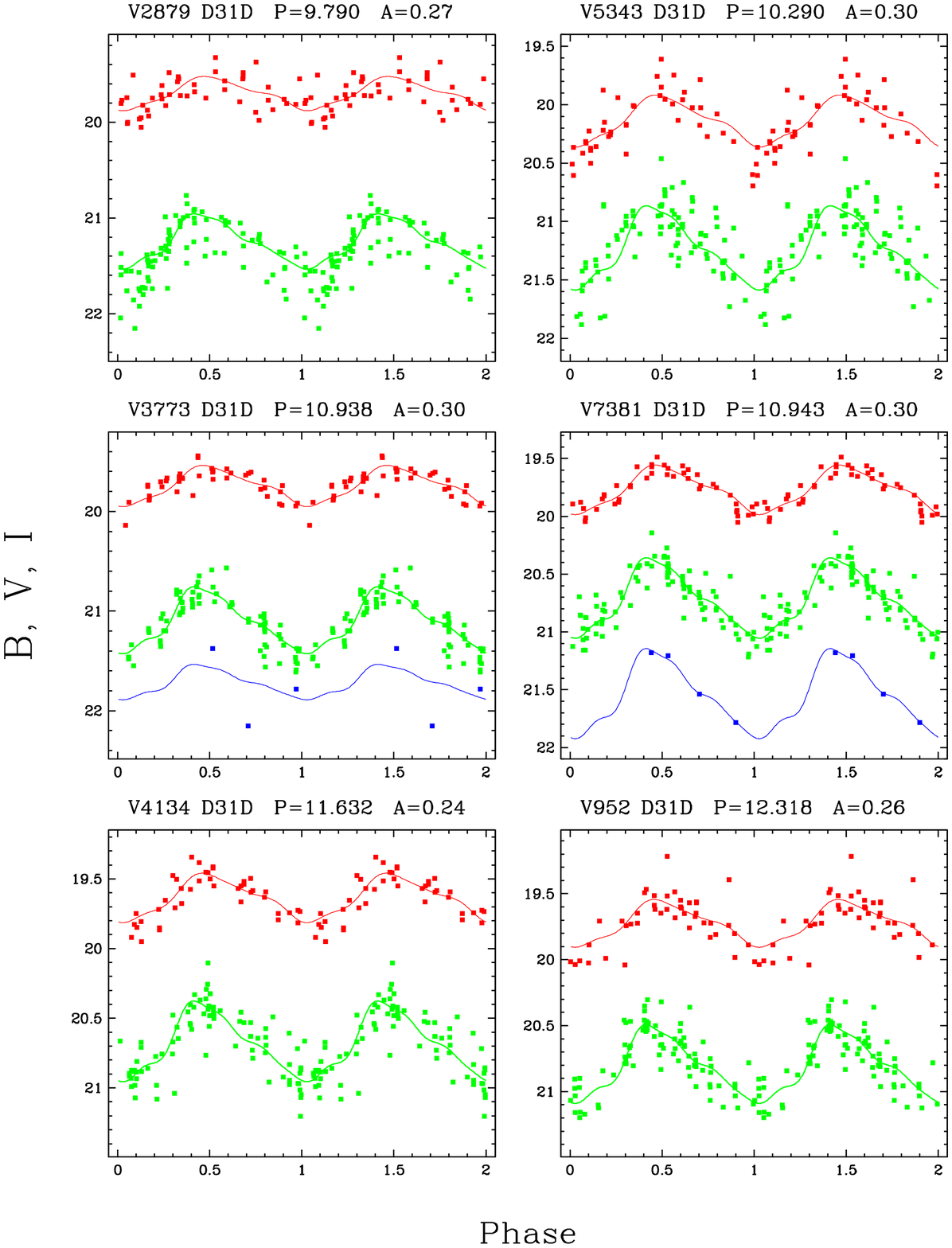}{19.5cm}{0}{83}{83}{-260}{-40}
\caption{Continued.}
\end{figure}

\addtocounter{figure}{-1}
\begin{figure}[p]
\plotfiddle{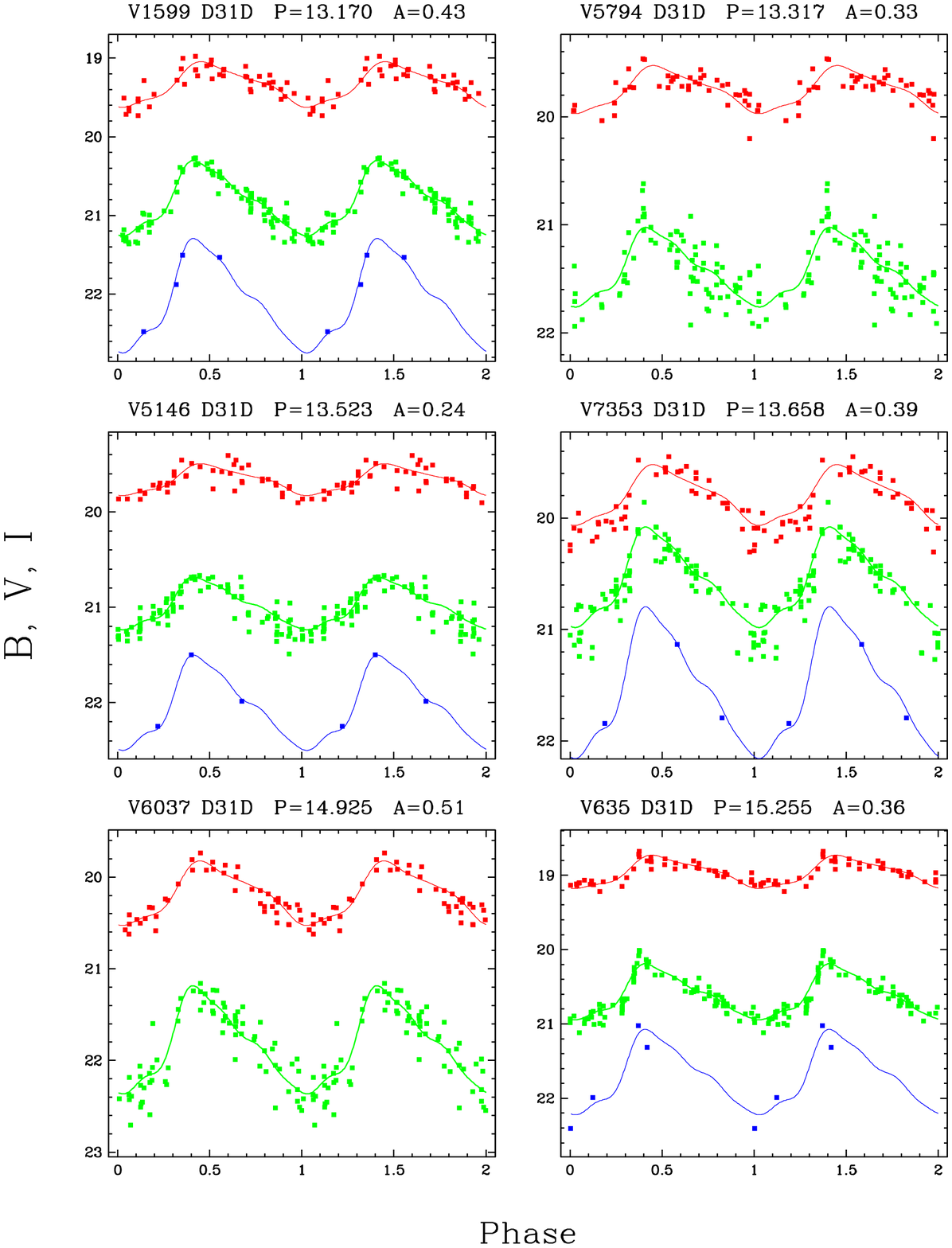}{19.5cm}{0}{83}{83}{-260}{-40}
\caption{Continued.}
\end{figure}

\addtocounter{figure}{-1}
\begin{figure}[p]
\plotfiddle{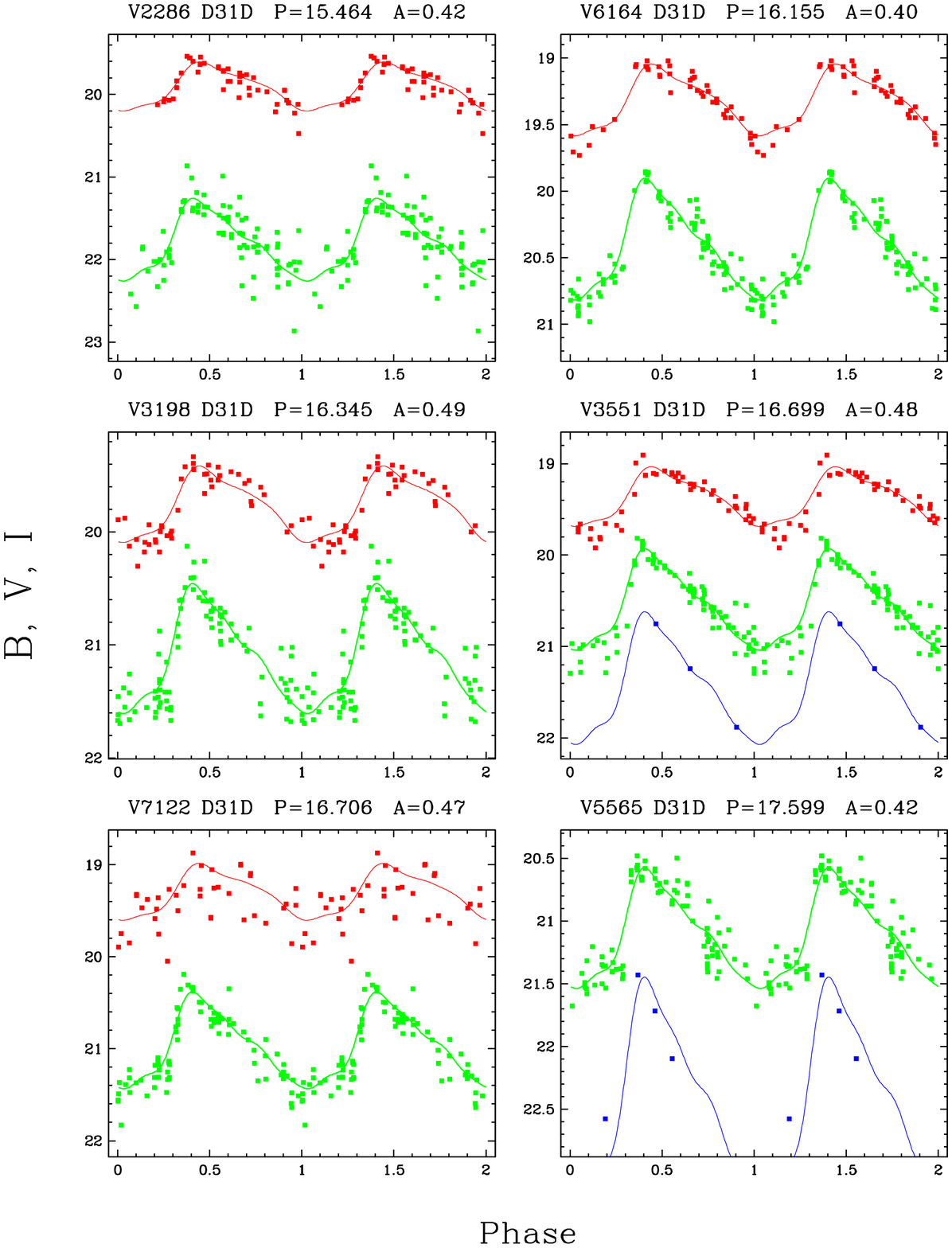}{19.5cm}{0}{83}{83}{-260}{-40}
\caption{Continued.}
\end{figure}

\addtocounter{figure}{-1}
\begin{figure}[p]
\plotfiddle{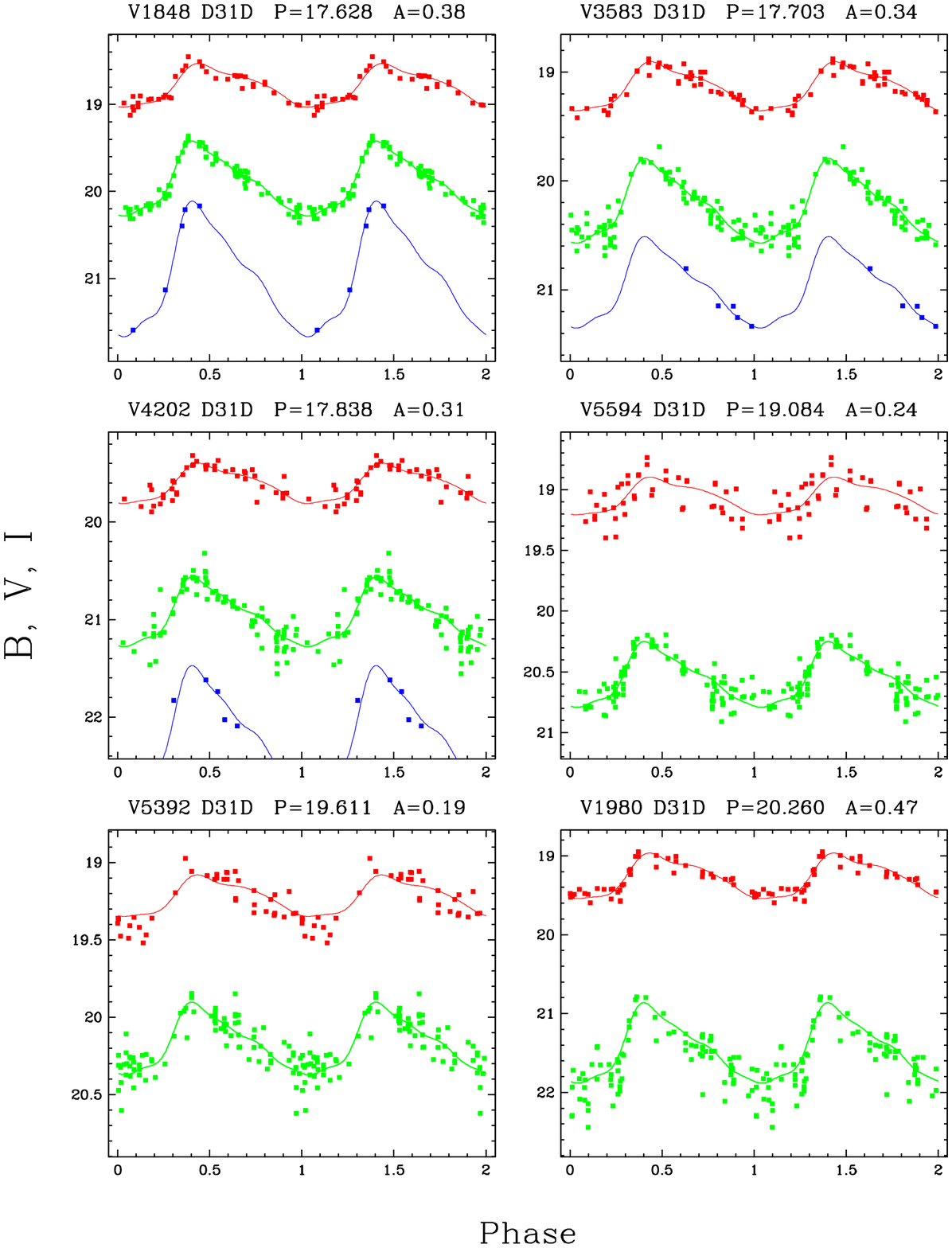}{19.5cm}{0}{83}{83}{-260}{-40}
\caption{Continued.}
\end{figure}

\addtocounter{figure}{-1}
\begin{figure}[p]
\plotfiddle{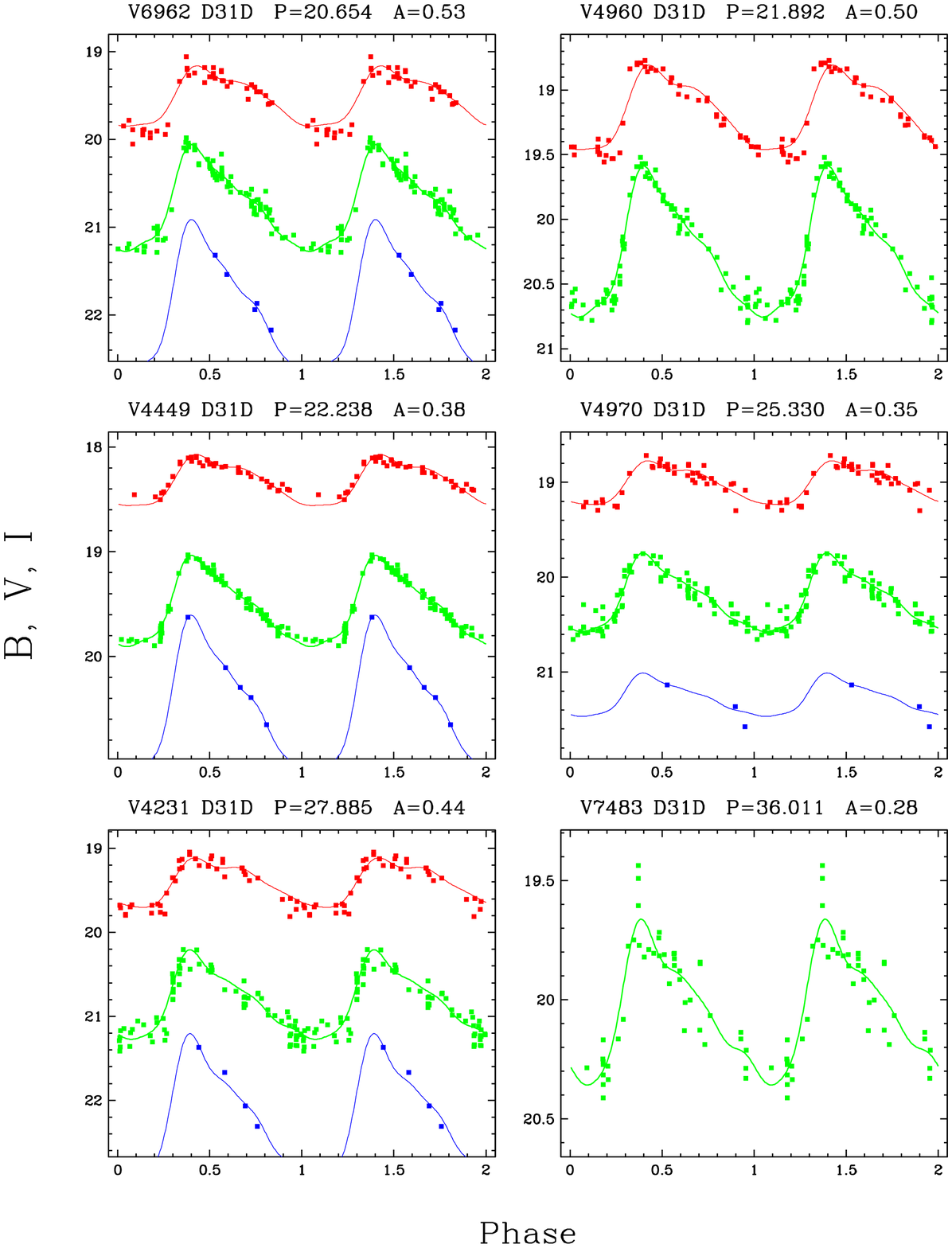}{19.5cm}{0}{83}{83}{-260}{-40}
\caption{Continued.}
\end{figure}

\addtocounter{figure}{-1}
\begin{figure}[p]
\plotfiddle{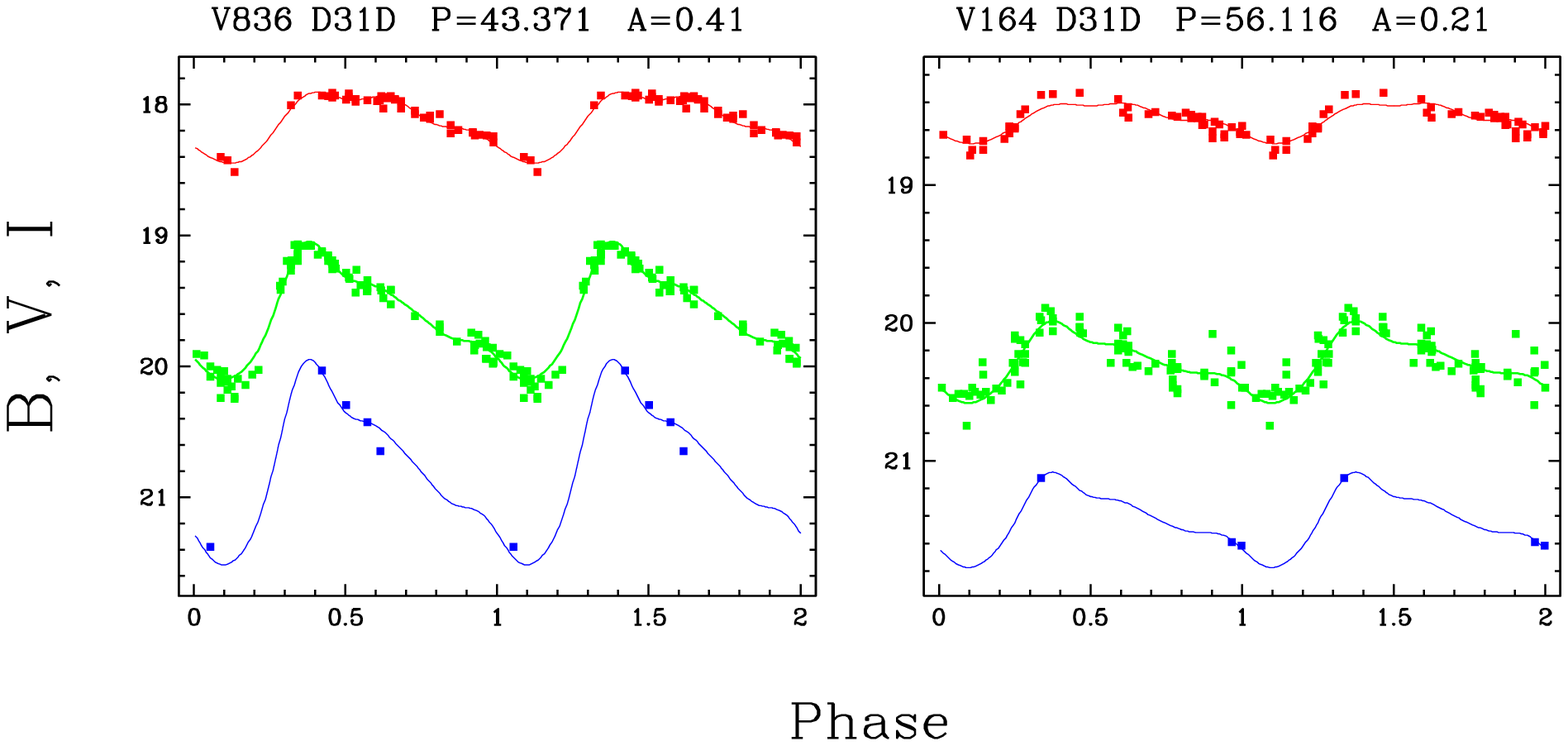}{6cm}{0}{83}{83}{-260}{-418}
\caption{Continued.}
\end{figure}

\subsection{Other periodic variables}

For many of the variables preliminarily classified as Cepheids we
decided upon closer examination to classify them as ``other periodic
variables''.  In Table~\ref{table:per} we present the parameters of 15
possible periodic variables, other than Cepheids and eclipsing
binaries, in the M31D field, sorted by the increasing period $P$.  For
each variable we present its name, J2000.0 coordinates, period $P$,
error-weighted mean magnitudes $\bar{V}$ and (when available)
$\bar{I}, \bar{B}$. To quantify the amplitude of the variability, we
also give the standard deviations of the measurements in the $BVI$
bands, $\sigma_{V},\sigma_{I}$ and $\sigma_{B}$.

Note that in most cases the periods were derived by fitting the
template Cepheids lightcurves, so they should only be treated as the
first approximation of the true period. Many of these periodic
variables are Type II Cepheids (W Virginis and RV Tauri variables),
based on their light curves and their location on the P-L diagram
(Figure~\ref{fig:pl}). One of the variables, V3994 D31D, as its location in
the CMD suggests, is probably an eclipsing binary.

\begin{figure}[p]
\plotfiddle{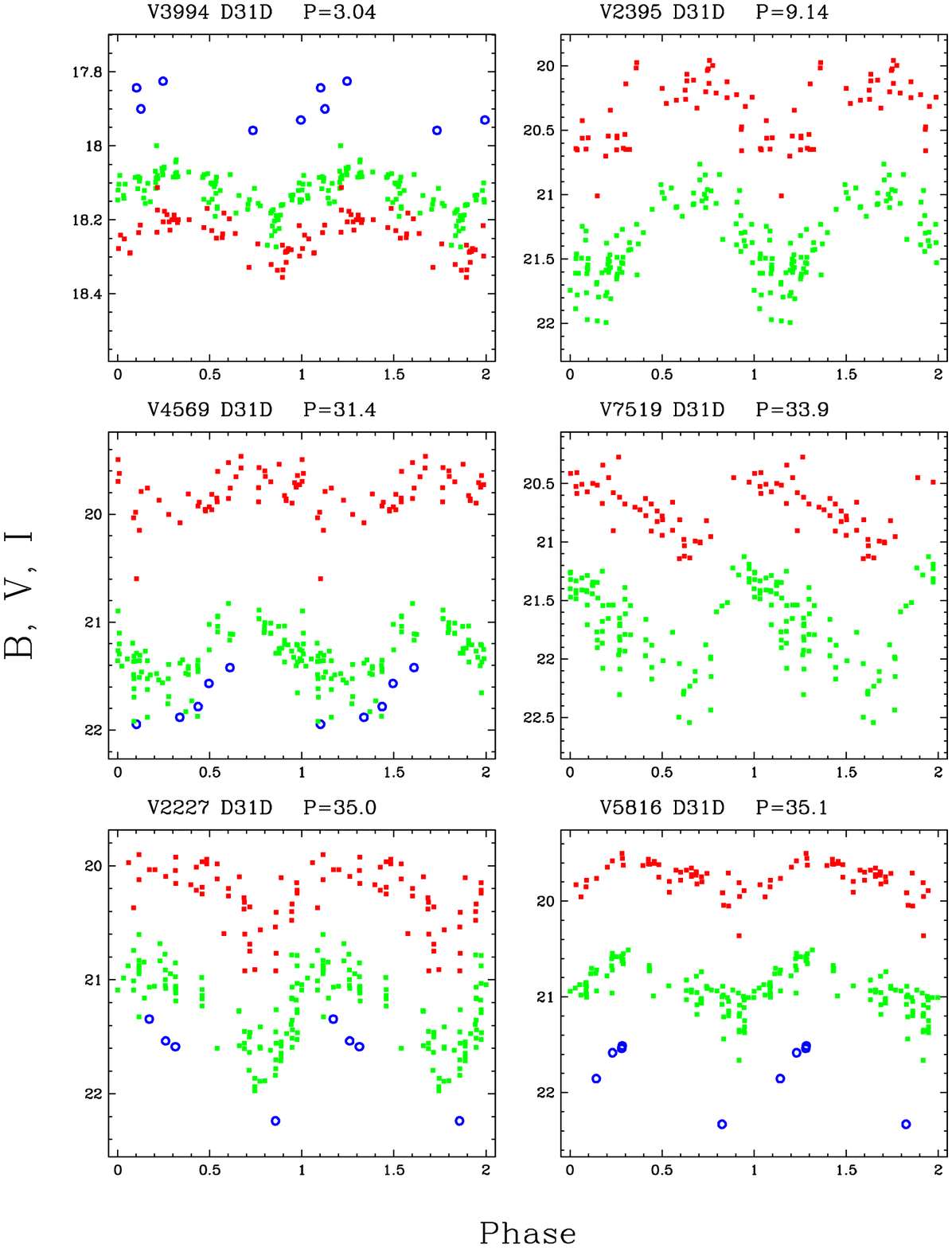}{19.5cm}{0}{83}{83}{-260}{-40}
\caption{$BVI$ lightcurves of other periodic variables found in the
field M31D.  $B$-band data (shown with the open circles, if present)
is usually the faintest and $I$ (if present) is usually the
brightest.}
\label{fig:per}
\end{figure}

\addtocounter{figure}{-1}
\begin{figure}[p]
\plotfiddle{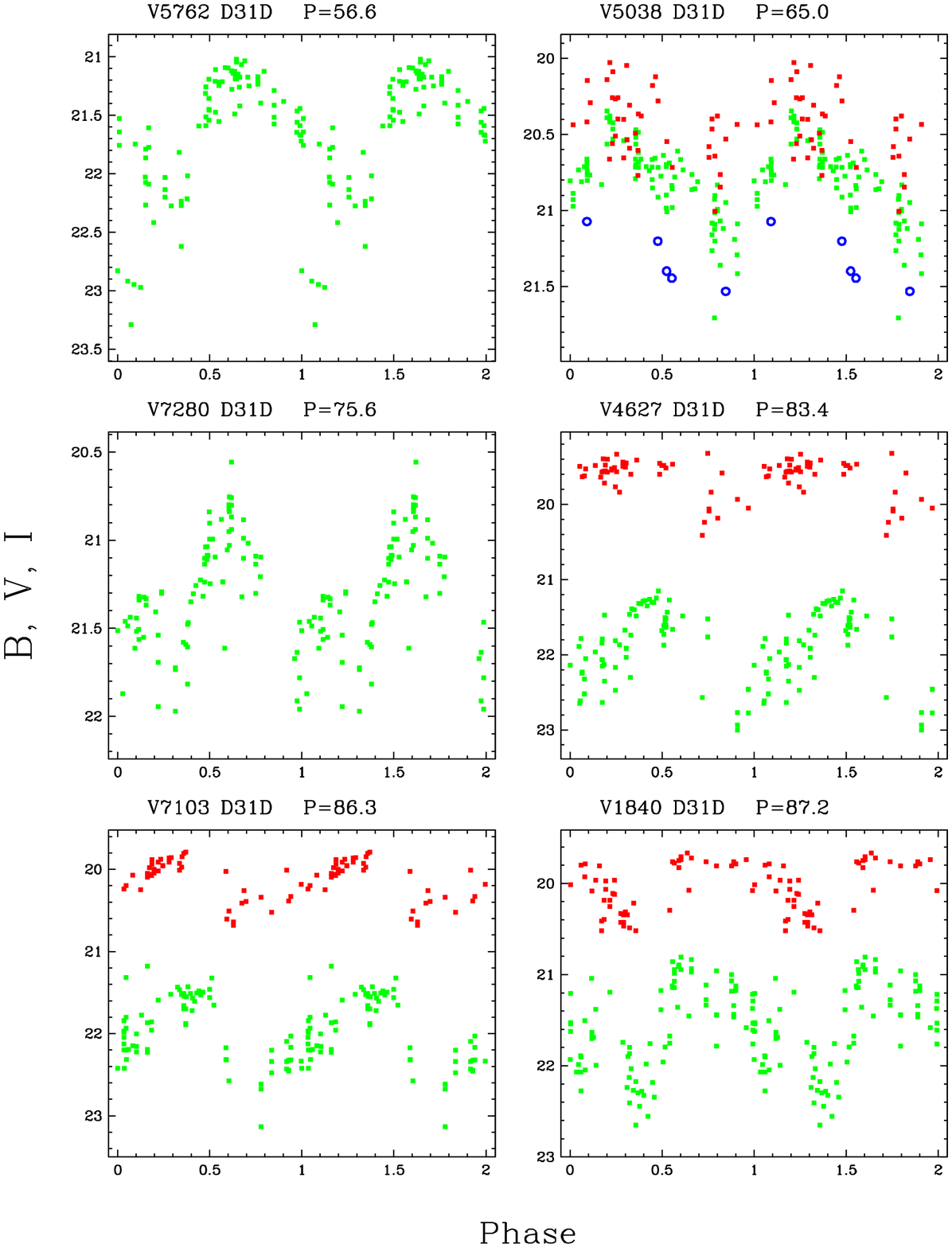}{19.5cm}{0}{83}{83}{-260}{-40}
\caption{Continued.}
\end{figure}

\addtocounter{figure}{-1}
\begin{figure}[p]
\plotfiddle{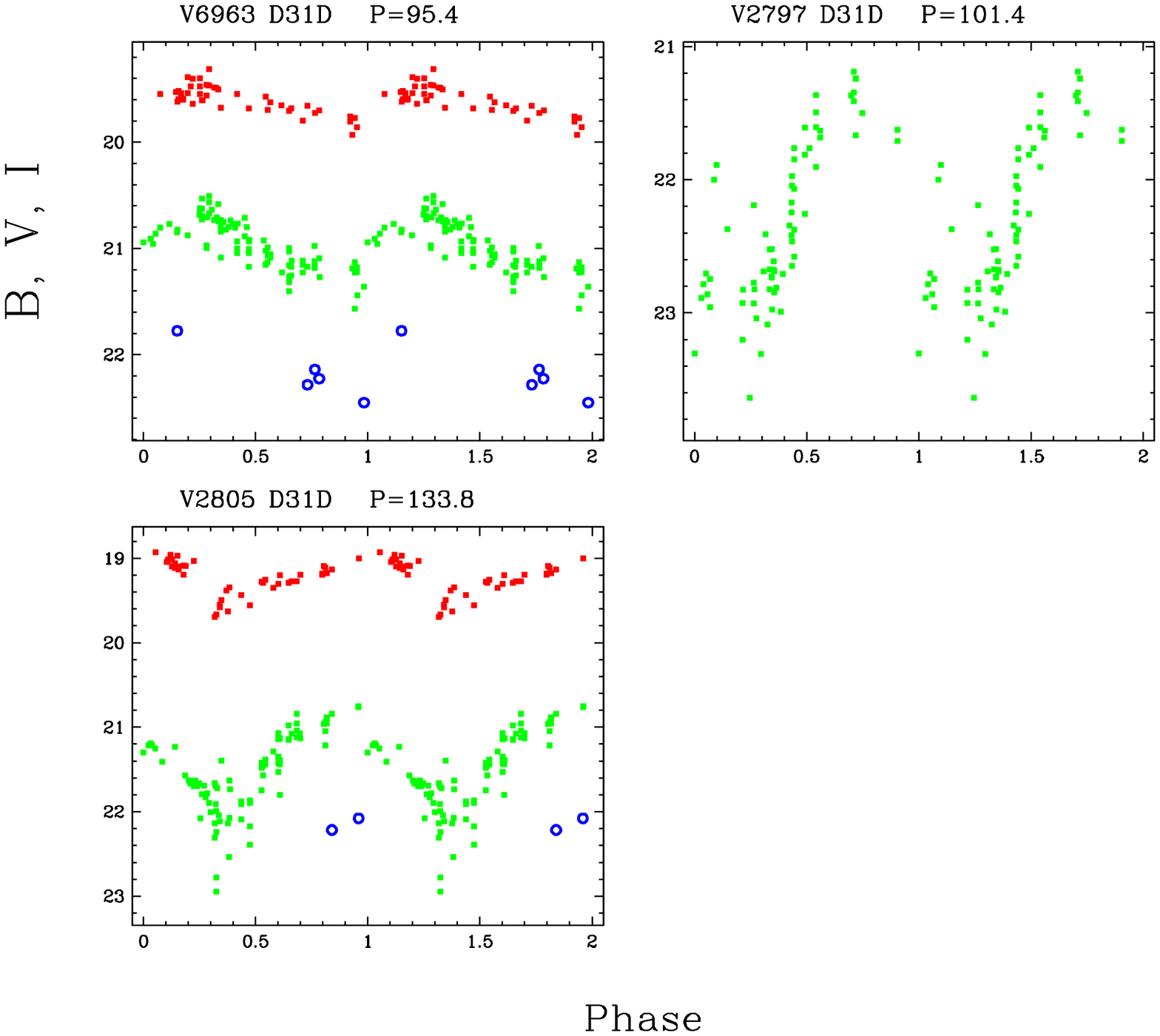}{9.25cm}{0}{83}{83}{-260}{-185}
\caption{Continued.}
\end{figure}

\begin{figure}[t]
\plotfiddle{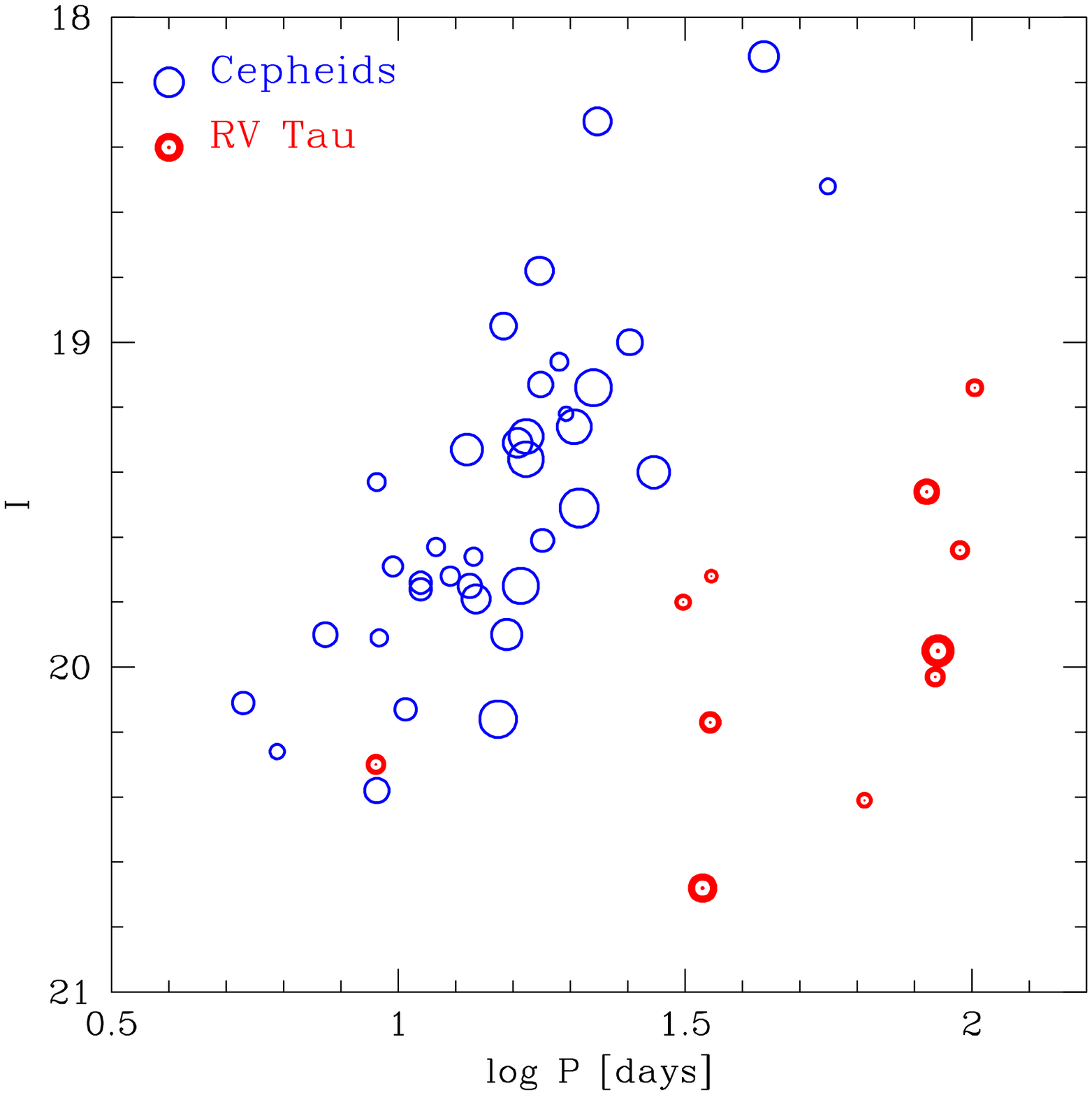}{8cm}{0}{50}{50}{-160}{-85}
\caption{Diagram of $\log{P}$ vs. $I$  for the
Cepheids (open circles) and RV Tau (dotted circles) variables. The
sizes of the circles are proportional to the $V$ amplitude of the
variability.}
\label{fig:pl}
\end{figure}

\begin{small}
\tablenum{2}
\begin{planotable}{lrrrrrrrl}
\tablewidth{35pc}
\tablecaption{DIRECT Cepheids in M31D}
\tablehead{ \colhead{Name} & \colhead{$\alpha_{J2000.0}$} &
\colhead{$\delta_{J2000.0}$} & \colhead{$P$}  &
\colhead{} & \colhead{} & \colhead{} & \colhead{} & \colhead{} \\ 
\colhead{(D31D)} & \colhead{(deg)} & \colhead{(deg)}
& \colhead{$(days)$} & \colhead{$\langle V\rangle$} & \colhead{$\langle I\rangle$} 
& \colhead{$\langle B\rangle$} & \colhead{$A$}  & \colhead{Comments} } 
\startdata 
 V7706\ldots   & 11.1179 & 41.3130 &  5.365 & 20.99 & 20.11 & 21.73 & 0.30 & Ma97 82  \nl 
 V7671\dotfill & 11.1144 & 41.3341 &  6.149 & 20.83 & 20.26 & 21.48 & 0.20 & Ma97 81  \nl 
  V883\dotfill & 10.9315 & 41.1969 &  7.459 & 20.92 & 19.90 & 21.57 & 0.33 &  \nl 
 V6311\dotfill & 11.0487 & 41.2732 &  9.172 & 21.69 & 20.38 &\nodata& 0.34 &  \nl 
 V1219\dotfill & 10.9364 & 41.2504 &  9.173 & 21.09 & 19.43 &\nodata& 0.24 &   \nl
 V4503\dotfill & 11.0004 & 41.2835 &  9.261 & 20.71 & 19.91 & 21.54 & 0.23 & Ma97 71  \nl 
 V2879\dotfill & 10.9716 & 41.2127 &  9.790 & 21.24 & 19.69 &\nodata& 0.27 &  \nl 
 V5343\dotfill & 11.0218 & 41.2345 & 10.290 & 21.22 & 20.13 &\nodata& 0.30 &  \nl 
 V3773\dotfill & 10.9875 & 41.2405 & 10.938 & 21.09 & 19.74 & 21.72 & 0.30 &  \nl 
 V7381\dotfill & 11.0925 & 41.3213 & 10.943 & 20.70 & 19.76 & 21.53 & 0.30 & Ma97 78  \nl 
 V4134\dotfill & 10.9927 & 41.2956 & 11.632 & 20.67 & 19.63 &\nodata& 0.24 &  \nl 
  V952\dotfill & 10.9332 & 41.1849 & 12.318 & 20.80 & 19.72 &\nodata& 0.26 &  \nl 
 V1599\dotfill & 10.9419 & 41.3027 & 13.170 & 20.78 & 19.33 & 21.98 & 0.43 &  \nl 
 V5794\dotfill & 11.0322 & 41.2944 & 13.317 & 21.40 & 19.75 &\nodata& 0.33 &  \nl 
 V5146\dotfill & 11.0166 & 41.2315 & 13.523 & 20.97 & 19.66 & 22.00 & 0.24 &  \nl 
 V7353\dotfill & 11.0908 & 41.3220 & 13.658 & 20.54 & 19.79 & 21.45 & 0.39 &  \nl 
 V6037\dotfill & 11.0401 & 41.3103 & 14.925 & 21.77 & 20.16 &\nodata& 0.51 &  \nl 
  V635\dotfill & 10.9252 & 41.2489 & 15.255 & 20.58 & 18.95 & 21.65 & 0.36 &  \nl 
 V2286\dotfill & 10.9576 & 41.2227 & 15.464 & 21.77 & 19.90 &\nodata& 0.42 &  \nl 
 V6164\dotfill & 11.0443 & 41.2838 & 16.155 & 20.38 & 19.31 &\nodata& 0.40 & Ma97 73  \nl 
 V3198\dotfill & 10.9772 & 41.2348 & 16.345 & 21.04 & 19.75 &\nodata& 0.49 &  \nl
 V3551\dotfill & 10.9835 & 41.2373 & 16.699 & 20.49 & 19.36 & 21.33 & 0.48 &  \nl 
 V7122\dotfill & 11.0783 & 41.3468 & 16.706 & 20.92 & 19.29 &\nodata& 0.47 & V7557 D31C \nl 
 V5565\dotfill & 11.0238 & 41.3485 & 17.599 & 21.07 &\nodata& 22.29 & 0.42 & V3277 D31C \nl 
 V1848\dotfill & 10.9455 & 41.3269 & 17.628 & 19.87 & 18.78 & 20.88 & 0.38 &  \nl 
 V3583\dotfill & 10.9849 & 41.2176 & 17.703 & 20.21 & 19.13 & 20.95 & 0.34 &  \nl 
 V4202\dotfill & 10.9952 & 41.2509 & 17.838 & 20.95 & 19.61 & 22.13 & 0.31 &  \nl 
 V5594\dotfill & 11.0272 & 41.2675 & 19.084 & 20.55 & 19.06 &\nodata& 0.24 &  \nl 
 V5392\dotfill & 11.0209 & 41.3162 & 19.611 & 20.16 & 19.22 &\nodata& 0.19 &  \nl 
 V1980\dotfill & 10.9496 & 41.2656 & 20.260 & 21.40 & 19.26 &\nodata& 0.47 &  \nl 
 V6962\dotfill & 11.0732 & 41.2807 & 20.654 & 20.69 & 19.51 & 21.79 & 0.53 & Ma97 74  \nl 
 V4960\dotfill & 11.0106 & 41.2787 & 21.892 & 20.20 & 19.14 &\nodata& 0.50 & Ma97 72  \nl
 V4449\dotfill & 10.9984 & 41.3149 & 22.238 & 19.50 & 18.32 & 20.38 & 0.38 & Ma97 70  \nl 
 V4970\dotfill & 11.0119 & 41.2402 & 25.330 & 20.19 & 19.00 & 21.26 & 0.35 &  \nl 
 V4231\dotfill & 10.9960 & 41.2386 & 27.885 & 20.77 & 19.40 & 22.07 & 0.44 &  \nl 
 V7483\dotfill & 11.0995 & 41.3533 & 36.011 & 20.03 &\nodata&\nodata& 0.28 & V9029 D31C \nl
  V836\dotfill & 10.9290 & 41.2476 & 43.371 & 19.57 & 18.12 & 20.69 & 0.41 &  \nl 
  V164\dotfill & 10.9177 & 41.1856 & 56.116 & 20.27 & 18.52 & 21.42 & 0.21 &  \nl 
\enddata
\label{table:ceph}
\tablecomments{V7122 D31D (Ma97 75) was found in paper III as V7557 D31C,
with $P=16.726\;days$, $\langle V\rangle = 20.85$, $\langle I\rangle = 
19.14$ and $\langle B\rangle = 21.98$; V5565 D31D was found as V3277 D31C, 
with $P=17.599\;days$, $\langle V\rangle = 21.02$, $\langle I\rangle =
19.60$ and $\langle B\rangle = 22.29$, V7483 D31D was found in paper III as
V9029 D31C, with $P=35.861\;days$, $\langle V\rangle = 20.31$, $\langle
I\rangle = 18.62$ and $\langle B\rangle = 21.39$.}
\end{planotable}
\end{small}

\subsection{Miscellaneous  variables}	
	
In Table~\ref{table:misc} we present the parameters of 13
miscellaneous variables in the M31D field, sorted by increasing value
of the mean magnitude $\bar{V}$. For each variable we present its
name, J2000.0 coordinates and mean magnitudes $\bar{V}, \bar{I}$ and
$\bar{B}$.  To quantify the amplitude of the variability, we also give
the standard deviations of the measurements in $VIB$ bands,
$\sigma_{V}, \sigma_{I}$ and $\sigma_{B}$.  In the ``Comments'' column
we give a rather broad sub-classification of the variability: LP --
possible long-period variable; Irr -- irregular variable.  In
Figure~\ref{fig:misc} we show the unphased $VI$ lightcurves of the
miscellaneous variables.

Many of the miscellaneous variables seem to represent the LP type of
variability. V7086 D31D is probably a Luminous Blue Variable. However,
inspection of the color-magnitude diagram (Figure~\ref{fig:cmd})
reveals that some of the miscellaneous variables land in the CMD in
the same area as the RV Tau variables, which suggests they are Type II
Cepheids.

\begin{small}
\tablenum{3} 
\begin{planotable}{cccrccccccl}
\tablewidth{40pc}
\tablecaption{DIRECT Other Periodic Variables in M31D}
\tablehead{ \colhead{Name} & \colhead{$\alpha_{J2000.0}$} &
\colhead{$\delta_{J2000.0}$} & \colhead{$P$} &
\colhead{} & \colhead{} & \colhead{} & \colhead{} & \colhead{} & \colhead{} & \colhead{} \\
\colhead{(D31D)} &  \colhead{(deg)} &  \colhead{(deg)} & 
\colhead{$(days)$} & \colhead{$\bar{V}$} &
\colhead{$\bar{I}$} & \colhead{$\bar{B}$} & \colhead{$\sigma_V$} & 
\colhead{$\sigma_I$} & \colhead{$\sigma_B$} & \colhead{Comments} }
\startdata     
 V3994\dotfill & 10.9898 & 41.2932 &   3.0 & 18.12 & 18.24 & 17.90 & 0.05 & 0.05 & 0.06 & EB? \nl 
 V2395\dotfill & 10.9578 & 41.2977 &   9.1 & 21.23 & 20.26 &\nodata& 0.29 & 0.25 &\nodata& W Virginis?\nl 
 V4569\dotfill & 11.0043 & 41.1923 &  31.4 & 21.25 & 19.75 & 21.66 & 0.24 & 0.20 & 0.22 & \nl 
 V7519\dotfill & 11.1053 & 41.2513 &  33.9 & 21.44 & 20.68 &\nodata& 0.35 & 0.23 &\nodata& \nl 
 V2227\dotfill & 10.9565 & 41.2106 &  35.0 & 21.12 & 20.17 & 21.57 & 0.33 & 0.28 & 0.39 & RV Tauri\nl 
 V5816\dotfill & 11.0352 & 41.2298 &  35.1 & 20.82 & 19.68 & 21.64 & 0.23 & 0.16 & 0.35 & \nl 
 V5762\dotfill & 11.0333 & 41.2120 &  56.6 & 21.30 &\nodata&\nodata& 0.51 &\nodata&\nodata& \nl 
 V5038\dotfill & 11.0117 & 41.3040 &  65.0 & 20.69 & 20.37 & 21.28 & 0.25 & 0.23 & 0.19 & \nl 
 V7280\dotfill & 11.0851 & 41.3495 &  75.6 & 21.16 &\nodata&\nodata& 0.33 &\nodata&\nodata& V8038 D31C \nl 
 V4627\dotfill & 11.0012 & 41.3400 &  83.4 & 21.50 & 19.53 &\nodata& 0.48 & 0.26 &\nodata& V1296 D31C \nl 
 V7103\dotfill & 11.0823 & 41.1845 &  86.3 & 21.68 & 20.02 &\nodata& 0.40 & 0.24 &\nodata& \nl 
 V1840\dotfill & 10.9453 & 41.3217 &  87.2 & 21.46 & 19.96 &\nodata& 0.47 & 0.27 &\nodata& RV Tauri\nl 
 V6963\dotfill & 11.0737 & 41.2648 &  95.4 & 20.86 & 19.58 & 22.10 & 0.23 & 0.13 & 0.25 & \nl 
 V2797\dotfill & 10.9687 & 41.2493 & 101.4 & 22.03 &\nodata&\nodata& 0.61 &\nodata&\nodata& \nl 
 V2805\dotfill & 10.9676 & 41.2914 & 133.8 & 21.28 & 19.15 & 22.16 & 0.46 & 0.20 & 0.10 & \nl 
\enddata
\label{table:per}
\tablecomments{Variables V7280 and V4627 were also found in Paper III.}
\end{planotable}
\end{small}

\begin{small}
\tablenum{4} 
\begin{planotable}{llllllllll}
\tablewidth{35pc}
\tablecaption{DIRECT Miscellaneous Variables in M31D}
\tablehead{ \colhead{Name} & \colhead{$\alpha_{J2000.0}$} &
\colhead{$\delta_{J2000.0}$} & \colhead{} & \colhead{} &
\colhead{} & \colhead{} & \colhead{} & \colhead{} & \colhead{} \\
\colhead{(D31D)} & \colhead{(deg)} & \colhead{(deg)} &
\colhead{$\bar{V}$} & \colhead{$\bar{I}$} & \colhead{$\bar{B}$} &
\colhead{$\sigma_V$} & \colhead{$\sigma_I$} & \colhead{$\sigma_B$} & \colhead{Comments} }
\startdata
 V7086\dotfill & 11.0768 & 41.3285 & 17.27 & 17.16 & 17.31 & 0.05 & 0.04 & 0.06 &  LBV\nl 
 V6157\dotfill & 11.0451 & 41.2538 & 19.52 & 18.02 &\nodata& 0.99 & 0.50 &\nodata&  \nl 
 V5592\dotfill & 11.0274 & 41.2601 & 19.52 & 16.72 & 21.55 & 0.17 & 0.08 & 0.08 &  LP \nl 
  V776\dotfill & 10.9279 & 41.2452 & 19.61 & 16.77 &\nodata& 0.12 & 0.05 &\nodata& LP \nl 
 V7855\dotfill & 11.1324 & 41.3190 & 19.87 & 17.18 & 21.72 & 0.22 & 0.13 & 0.06 &  LP\nl 
 V3712\dotfill & 10.9878 & 41.1934 & 19.89 & 16.66 &\nodata& 0.27 & 0.12 &\nodata&  LP\nl 
 V2136\dotfill & 10.9546 & 41.2046 & 21.00 & 17.67 &\nodata& 0.54 & 0.92 &\nodata&  LP\nl 
 V5309\dotfill & 11.0175 & 41.3357 & 21.01 & 19.21 &\nodata& 0.79 & 0.43 &\nodata& V2679 D31C \nl 
 V6371\dotfill & 11.0498 & 41.3000 & 21.19 & 19.71 &\nodata& 0.82 & 0.68 &\nodata& RV Tauri \nl 
 V7949\dotfill & 11.1451 & 41.2299 & 21.21 & 19.05 &\nodata& 0.83 & 0.25 &\nodata& RV Tauri \nl 
 V7725\dotfill & 11.1228 & 41.2280 & 21.47 & 19.85 &\nodata& 0.84 & 0.42 &\nodata& RV Tauri \nl 
 V5323\dotfill & 11.0175 & 41.3497 & 21.47 &\nodata&\nodata& 0.56 &\nodata&\nodata& V2724 D31C \nl 
 V4599\dotfill & 11.0041 & 41.2206 & 21.54 & 19.34 &\nodata& 0.40 & 0.45 &\nodata& RV Tauri \nl 
\enddata
\label{table:misc}
\tablecomments{Variables V5309 and V5323 were also found in Paper III.}
\end{planotable}
\end{small}

\subsection{Comparison with other catalogs}

The area of M31D field has not been observed frequently before and the
only overlapping variable star catalog is given by Magnier et
al.~(1997, hereafter Ma97). Out of the nine variable stars in Ma97
which are in our M31D field, we cross-identified all nine, also
classifying them as Cepheids (see Table~\ref{table:ceph} for
cross-ids).

There was also by design a slight overlap between the M31D and M31C
fields (Figure~\ref{fig:xy}). There were three Cepheids from the M31D
field in the overlap region, and they were all cross-identified in the
M31C catalog with very similar properties of their light curves (see
Table~\ref{table:ceph}). One other Cepheid given in Paper III, V11190
D31C, failed to qualify as a variable candidate, with $J_S=0.294$. We
also cross-identified two other periodic variables found in the
overlap (see Table~\ref{table:per}).  Two miscellaneous variables in
the M31D field were cross-identified with periodic variables in the
M31C field (see Table~\ref{table:misc}). None of the five
miscellaneous variables detected in the overlapping part of the M31C
field were classified as variable stars in the M31D field. Four of
them had $N_{good}<45$. One turned out to have $\sigma=0.03$, thus
very narrowly failing to qualify as a variable.

\begin{figure}[p]
\plotfiddle{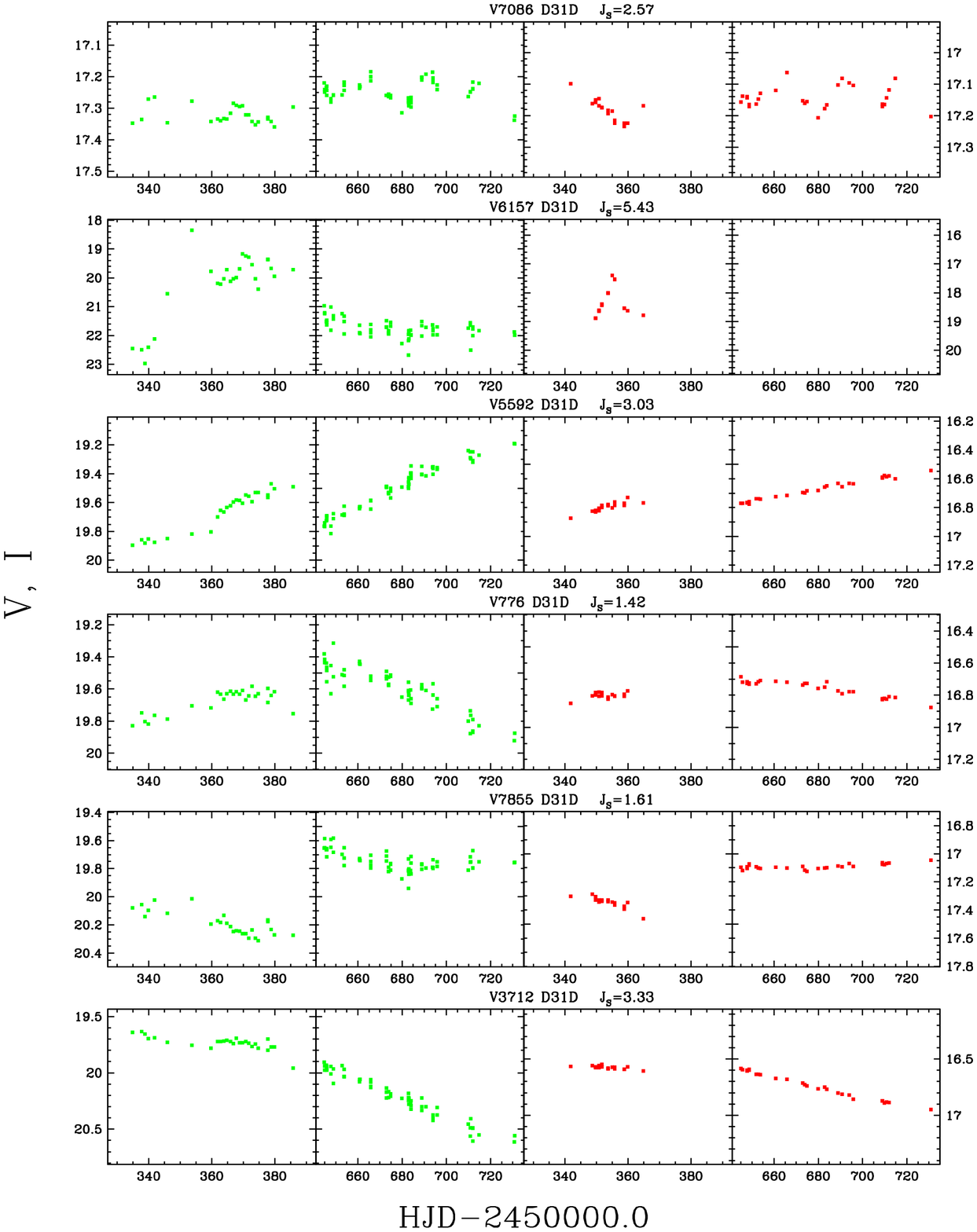}{19.5cm}{0}{83}{83}{-260}{-40}
\caption{$VI$ lightcurves of the miscellaneous variables found in the
field M31D.  $I$ (if present) is plotted in the two right panels.
$B$-band data is not shown.}
\label{fig:misc}
\end{figure}

\addtocounter{figure}{-1}
\begin{figure}[p]
\plotfiddle{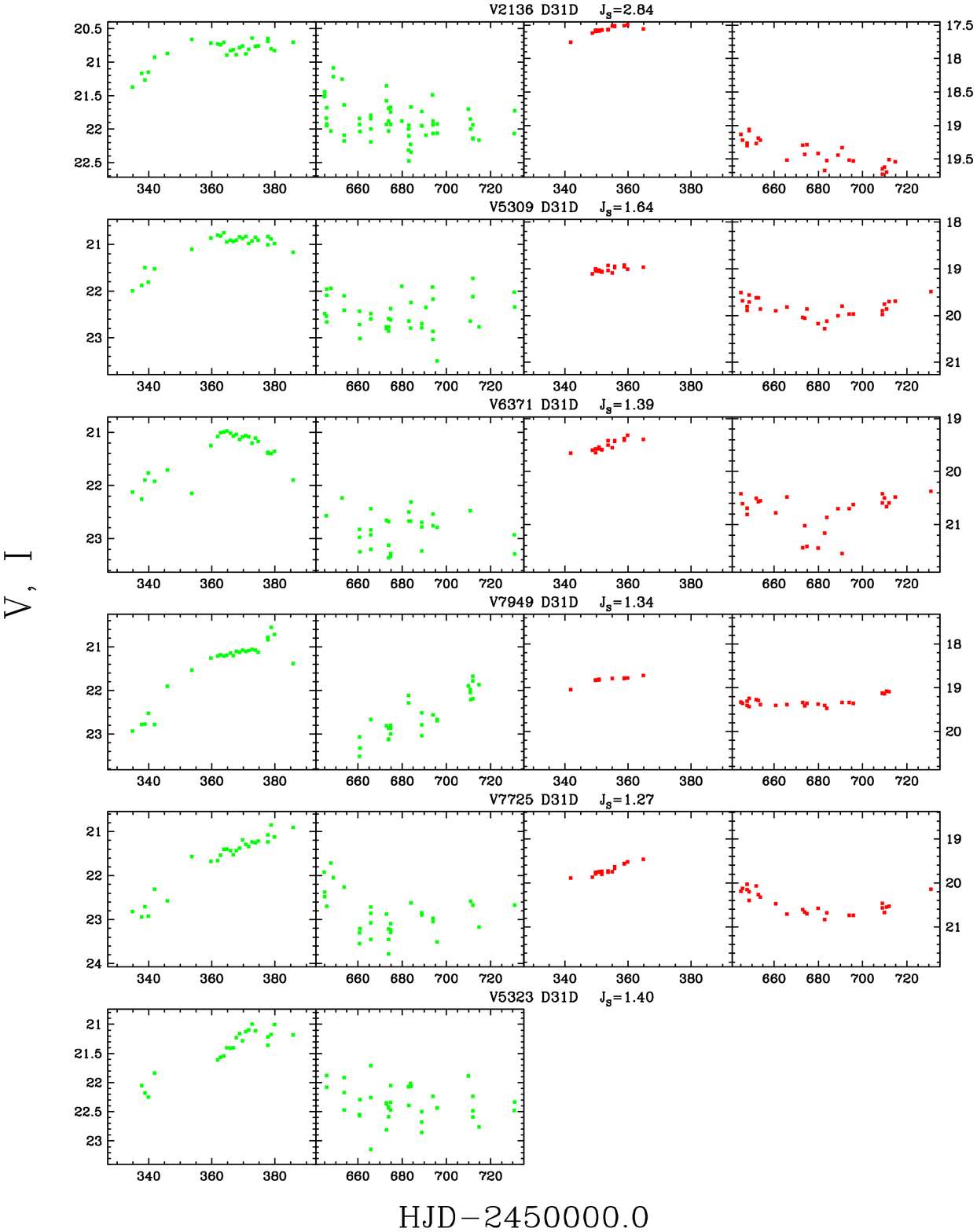}{19.5cm}{0}{83}{83}{-260}{-40}
\caption{Continued.}
\end{figure}

\addtocounter{figure}{-1}
\begin{figure}[p]
\plotfiddle{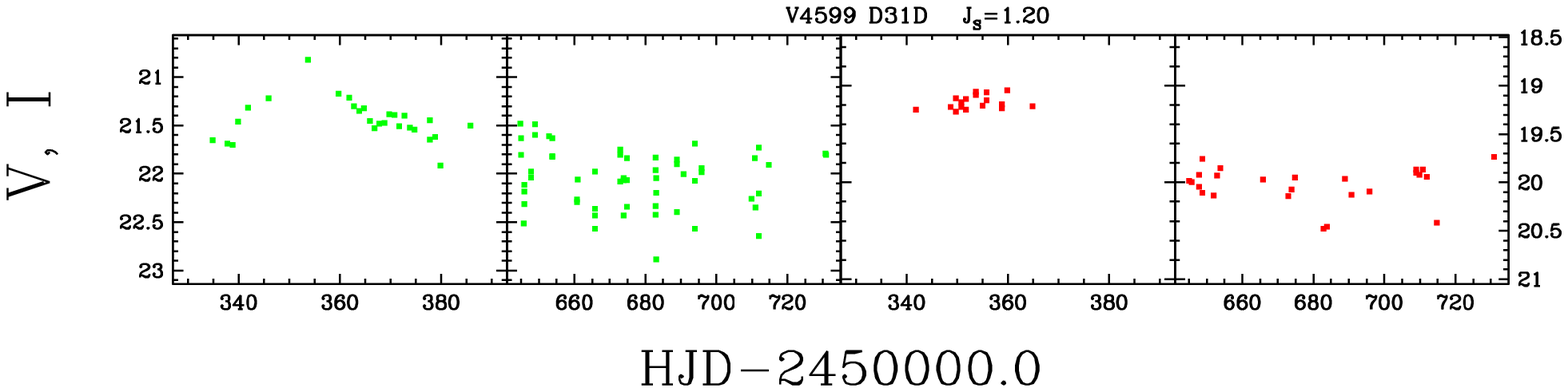}{2cm}{0}{83}{83}{-260}{-495}
\caption{Continued.}
\end{figure}
\newpage
\section{Discussion}

\begin{figure}[t]
\plotfiddle{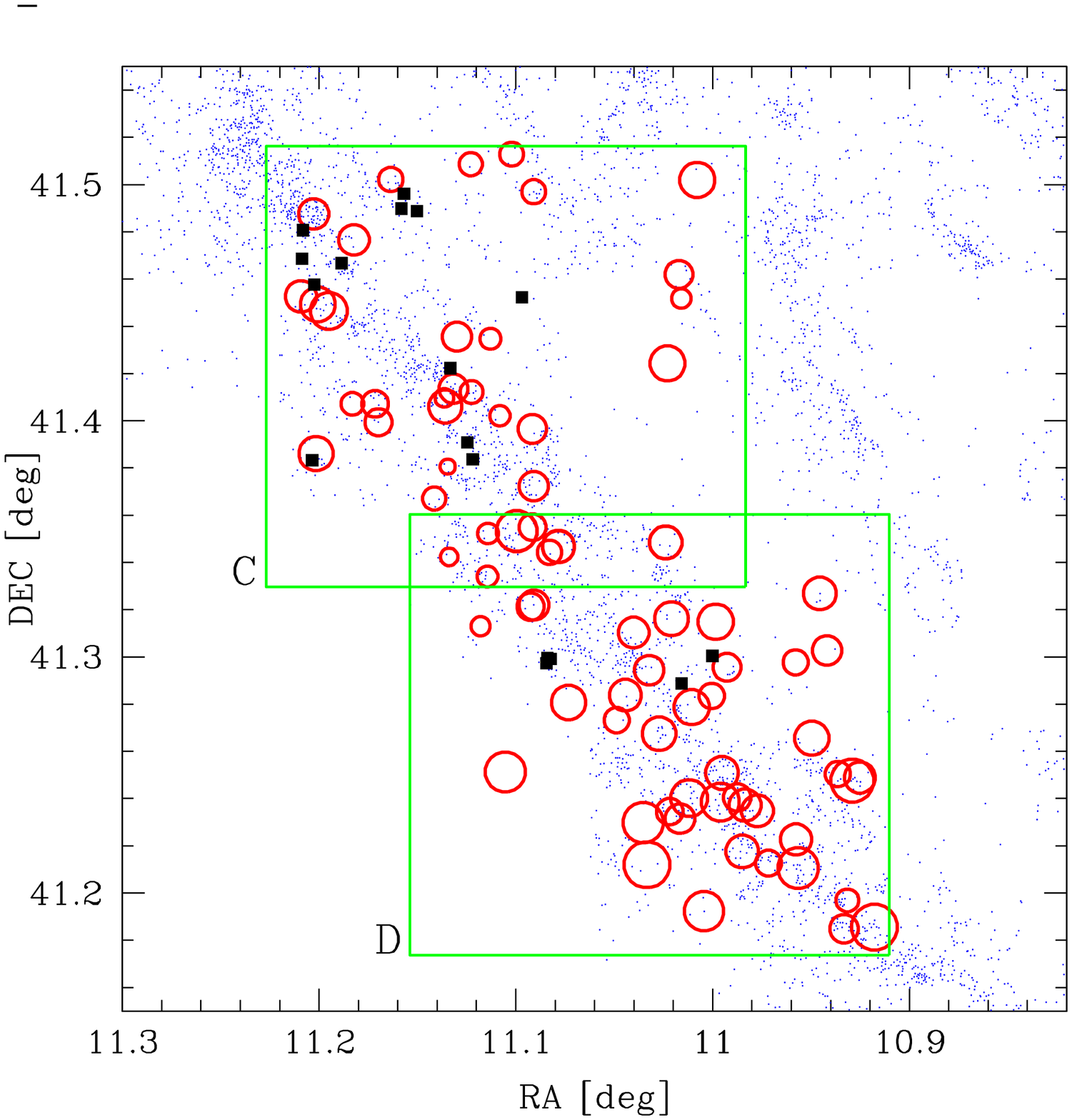}{8cm}{0}{50}{50}{-160}{-85}
\caption{Location of eclipsing binaries (filled squares) and Cepheids
(open circles) in the fields M31D and M31C, along with the blue stars
($B-V<0.4$) selected from the photometric survey of M31 by Magnier et
al.~(1992) and Haiman et al.~(1994). The sizes of the circles
representing the Cepheids variables are proportional to the logarithm
of their period.
\label{fig:xy}}
\end{figure}
 
In Figure~\ref{fig:cmd} we show $V,\;V-I$ and $V,\;B-V$
color-magnitude diagrams for the variable stars found in the field
M31D. The eclipsing binaries and Cepheids are plotted in the left
panels and the other periodic variables and miscellaneous variables
are plotted in the right panels. As expected, the eclipsing 
binaries occupy the blue upper main sequence of M31 stars. The Cepheid 
variables group near $B-V\sim1.0$, with considerable
scatter probably due to the differential reddening across the field. The
other periodic variable stars have positions on the CMD similar to the
Cepheids, with the exception of V3994 D31D, which is probably an eclipsing
binary. The miscellaneous variables are scattered throughout the
CMDs and represent several classes of variability. Many of them are
very red with $V-I>2.0$, and are probably Mira variables. The
brightest miscellaneous variable is probably a foreground star
belonging to our Galaxy. 

In Figure~\ref{fig:xy} we plot the location of eclipsing binaries and
Cepheids in the fields M31D and M31C, along with the blue stars
($B-V<0.4$) selected from the photometric survey of M31 by Magnier et
al.~(1992) and Haiman et al.~(1994). The sizes of the circles
representing the Cepheids variables are proportional to the logarithm
of their period. As could have been expected, both types of variables
group along the spiral arms, as they represent relatively young
populations of stars.  We will explore various properties of our
sample of Cepheids in the future paper (Sasselov et al.~1999, in
preparation).

\acknowledgments{We would like to thank the TAC of the
Michigan-Dartmouth-MIT (MDM) Observatory and the TAC of the
F.~L.~Whipple Observatory (FLWO) for the generous amounts of telescope
time devoted to this project. We are very grateful to Bohdan
Paczy\'nski for motivating us to undertake this project and his always
helpful comments and suggestions.  We thank Lucas Macri for taking
some of the data described in this paper and Przemek Wo\'zniak for his
FITS-manipulation programs. The staff of the MDM and the FLWO
observatories is thanked for their support during the long observing
runs.  KZS was supported by the Harvard-Smithsonian Center for
Astrophysics Fellowship. JK was supported by NSF grant AST-9528096 to
Bohdan Paczy\'nski and by the Polish KBN grant 2P03D011.12. JLT was
supported by the NSF grant AST-9401519.}

\end{document}